\documentclass[%
 reprint,
 superscriptaddress,
 amsmath,amssymb,
 aps,
 pra,
 nolongbibliography,
]{revtex4-2}

\usepackage{glossaries}
\usepackage[hidelinks]{hyperref}
\usepackage[capitalise]{cleveref}
\usepackage{graphicx}
\usepackage{dcolumn}
\usepackage{bm}
\usepackage{layouts}%
\usepackage{xcolor,soul}

\newcommand*\diff{\mathop{}\!\mathrm{d}}
\newcommand{\E}[1]{\times 10^{#1}}

\newacronym{pic}{PIC}{particle-in-cell}
\newacronym{pwfa}{PWFA}{plasma wakefield accelerator}
\newacronym{cfi}{CFI}{current filamentation instability}
\newacronym{tsi}{TSI}{(longitudinal) two-stream instability}
\newacronym{obi}{OBI}{oblique instability}
\newacronym{tts}{TTS}{transverse two-stream instability}
\newacronym{smi}{SMI}{self-modulation instability}

\newcommand{\im}{{i\mkern1mu}}
\newenvironment{nalign}{
    \begin{equation}
    \begin{aligned}
}{
    \end{aligned}
    \end{equation}
    \ignorespacesafterend
}
\newenvironment{ngather}{
    \begin{equation}
    \begin{gathered}
}{
    \end{gathered}
    \end{equation}
    \ignorespacesafterend
}

\begin{document}

\preprint{APS/123-QED}

\title{Wakefield-driven filamentation of warm beams in plasma}

\author{Erwin Walter}
\email{erwin.walter@ipp.mpg.de}
\affiliation{Max Planck Institute for Plasma Physics, 85748 Garching, Germany}
\affiliation{Exzellenzcluster ORIGINS, 85748 Garching, Germany}
\author{John P. Farmer}%
\email{j.farmer@cern.ch}
\affiliation{%
Max Planck Institute for Physics, 80805 Munich, Germany
}%
\author{Martin S. Weidl}%
\affiliation{Max Planck Institute for Plasma Physics, 85748 Garching, Germany}
\author{Alexander Pukhov}%
\affiliation{%
University of Duesseldorf, 40225 Duesseldorf, Germany
}%
\author{Frank Jenko}%
\affiliation{Max Planck Institute for Plasma Physics, 85748 Garching, Germany}

\date{\today}

\begin{abstract}
  Charged and quasi-neutral beams propagating through an unmagnetised plasma are subject to numerous collisionless instabilities on the small scale of the plasma skin depth. The electrostatic two-stream instability, driven by longitudinal and transverse wakefields, dominates for dilute beams. This leads to modulation of the beam along the propagation direction and, for wide beams, transverse filamentation. A three-dimensional spatiotemporal two-stream theory for warm beams with a finite extent is developed. Unlike the cold beam limit, diffusion due to a finite emittance gives rise to a dominant wavenumber, and a cut-off wavenumber above which filamentation is suppressed. Particle-in-cell simulations with quasineutral electron-positron beams in the relativistic regime give excellent agreement with the theoretical model. This work provides deeper insights into the effect of diffusion on filamentation of finite beams, crucial for comprehending plasma-based accelerators in laboratory and cosmic settings.
\end{abstract}

\maketitle

\section{Introduction}

From supernovae in distant galaxies to laboratory-based wakefield accelerators, the collisionless interaction of relativistic particles with plasma is relevant to many physical scales. The interactions are often governed by kinetic micro-instabilities, which result in electrostatic and electromagnetic fluctuations \citep{Chen2016, Bret2010, Michno2010}. This dissipation of a directed relativistic flow transfers kinetic energy to field energy, which can give rise to collisionless shocks in the astrophysical context. In these collisionless shocks, non-thermal particles accelerated to TeV energies through Fermi-type processes \citep{Spitkovsky2008} or Landau resonance \citep{Landau1946, Iwamoto2019, Tajima2020} emit synchrotron radiation across a spectrum from radio to gamma-ray frequencies \citep{Hillas1984, Piron2016}. Collisionless shocks are observed in active galactic nuclei and supernovae-remnants \citep{Bohdan2021}, or in gamma-ray bursts that occur during merge events of neutron stars or black holes \citep{Medvedev1999, Perna2016}.

Specially designed experimental setups \cite{Fiuza2020, Arrowsmith2021, Zhang2022} have recently enabled unprecedented investigations of electromagnetic plasma instabilities relevant on the astronomical scale.
Beam-driven \glspl{pwfa} \citep{Chen1985}, which can be utilised as $\gamma$-ray sources \citep{Macchi2018} or to achieve higher accelerating fields compared to conventional RF accelerators \citep{Schroeder2011, Caldwell2011, Adli2018}, are also subject to microinstabilities. Furthermore, \glspl{pwfa} can be adapted to investigate regimes relevant to astrophysics \citep{Verra2024, Allen2012}.

The interaction of a relativistic beam with an unmagnetised plasma can be usually categorised between the electromagnetic Weibel-like \gls{cfi} \citep{Weibel1959, Fried1959}, driven by the plasma return current, or two-stream instabilities \citep{Bludman1960, Fainberg1969, Bret2004}, driven by the electrostatic plasma response. In the latter, the beam excites Langmuir plasma waves \citep{Langmuir1929}, conventionally called wakefields in particle accelerators \citep{Dawson1959}, which lead to the \gls{tsi} and the \gls{tts} \citep{Lawson1977}. The combination of TSI and TTS is usually referred to as the \gls{obi} \citep{Watson1960, Bret2004} and allows dilute beams to undergo a similar filamentary behaviour as \gls{cfi}.

Previous theoretical work on \gls{cfi} for cold, spatially uniform streams determined that the temporal growth rate increases with transverse wavenumber \citep{Davidson1972}. These studies were extended to warm streams, in which diffusion acts to suppress small-scale filamentation, and a dominant wavenumber was calculated \citep{Silva2002, Jia2013}. For cold longitudinally bounded streams, \gls{cfi} was found to exhibit spatiotemporal growth at the beam head \citep{Pathak2015}.

For two-stream instabilities in cold uniform streams, the growth rate also increases with transverse wavenumber \citep{Watson1960, Bret2004}. It was predicted that diffusion would suppress the growth of small-scale filaments \citep{Bludman1960}, which was later studied numerically, and a threshold above which the system is stable was found analytically \citep{Bret2010b}. For a localised disturbance in cold bounded systems, \gls{tsi} \citep{Bers1983, Jones1983} and \gls{tts} \citep{Claveria2022} demonstrate a pulse-shaped spatiotemporal growth. However, the effect of a finite beam emittance on the spatiotemporal growth of the filamentation instability has not previously been treated analytically.

This manuscript introduces a fully three-dimensional, spatiotemporal theory describing filamentation of a warm beam due to wakefield-driven two-stream instabilities. This allows limits to be set on the beam temperature for laboratory astrophysics schemes seeking to investigate these instabilities and PWFA experiments seeking to avoid them. The work is structured as follows: Wakefield-driven filamentation is introduced in \cref{sec:wakefield_filamentation}. In \cref{sec:theory_wf_cold}, an analytical expression for the growth is derived for a cold bounded beam with a transverse profile. The theory is extended to warm beams in \cref{sec:theory_wf_warm}, which considers the effect of diffusion. This allows the exact value for the dominant wavenumber to be calculated, as well as the cut-off above which no filamentation occurs. The analytical predictions are throughout compared to two and three-dimensional \gls{pic} simulations, which show excellent agreement.

\section{\label{sec:wakefield_filamentation}Wakefield-Driven Filamentation}

The regimes for the two filamentation instabilities are defined by the current imbalance in the system. The beam and plasma currents must be comparable for \gls{cfi} to dominate. For a relativistic beam propagating in stationary plasma, relevant to many astrophysical schemes, this requires a dense beam, $n_b \gtrsim n_p$, with $n_b$ and $n_p$ the beam and plasma density \citep{Bret2010}.
For a dilute beam, $n_b\ll n_p$, the plasma current is negligible, and plasma electrons are mainly deflected by the beam charge. The resulting wakefield leads to \gls{tsi} and \gls{tts} \citep{Bret2010, Keinings1987, Katsouleas1987}.

Plasma wakefield experiments use a charged beam, which is usually dense and short, $k_p\sigma_\zeta<1$, with $\sigma_\zeta$ the rms length \citep{Clayton2016, Albert2021}. Here, $k_p=\omega_p/c$ is the plasma wavenumber, where $c$ is the speed of light and $\omega_p=[e^2 n_p/(\varepsilon_0 m_e)]^{1/2}$ is the plasma frequency, with $e$ the elementary charge, $m_e$ the electron mass, and $\varepsilon_0$ the vacuum permittivity. A dilute and long beam, $k_p\sigma_\zeta\gg 1$, is subject to \gls{tts}.
For narrow beams, $k_p\sigma_r\lesssim 1$, with $\sigma_r$ the rms width, \gls{tts} can take the form of the axisymmetric \gls{smi} modulating the beam radius \citep{Schroeder2011}, or the antisymmetric hosing instability displacing the beam centroid \citep{Moreira2023}.

Fully modulated, the beam can resonantly drive a quasi-linear wake with an accelerating field comparable to that driven by a short, dense beam \citep{Gschwendtner2022}. Wakefield experiments do not utilise wide beams as they may undergo filamentation due to transverse perturbations \citep{Verra2024} and degrade the wakefield. Experiments that investigate filamentation instabilities may operate with quasi-neutral beams (equal populations of particles with
opposite charge) to suppress \gls{smi} \citep{Shukla2018}.

\begin{figure}
  \includegraphics[width=\linewidth]{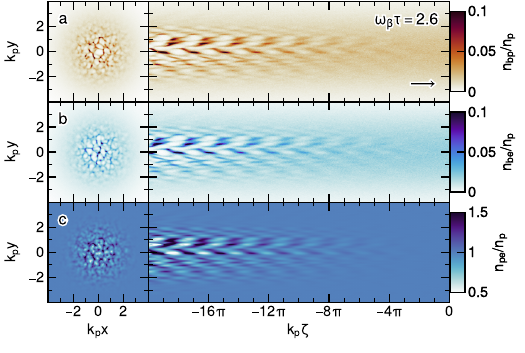}
  \caption{\label{fig:tts_3d_intro} Filamentation of a dilute bunch due to wakefields. Transverse and longitudinal slices of the a) positron and b) electron density of a dilute quasineutral bunch after propagating $\omega_\beta \tau=2.6$ in an initially uniform plasma. c) The electron density of the corresponding plasma response. The bunch propagates to the right, with its head at $\zeta=0$. The transverse and longitudinal slices are taken at $k_p\zeta=-18\pi$ and $k_p x=0$, respectively.}
\end{figure}

This filamentation of a quasi-neutral, dilute bunch and the corresponding plasma response is shown in \cref{fig:tts_3d_intro} after propagating $2.6/\omega_\beta$ in plasma, where $\omega_\beta=[q_b^2 n_b/(2\gamma_b \varepsilon_0 m_b)]^{1/2}$ is the betatron frequency, with $q_b$ the charge, $\gamma_b$ the Lorentz factor and $m_b$ the mass of the bunch particles. Both the bunch and the plasma response exhibit roughly equidistant filaments, where positrons and electrons are oppositely aligned due to the plasma wakefield that drives the instability. The plasma electrons align with the bunch positrons driven by the bunch charge. A periodic modulation occurs along the bunch, arising due to the oscillation of the wakefield.

The simulation in \cref{fig:tts_3d_intro} was carried out using the three-dimensional, quasistatic PIC code qv3d \citep{Pukhov2016}, built on the VLPL platform \citep{Pukhov1999}. The relativistic, $\gamma_b=22.4$ ($u_b/c=0.999$), warm electron-positron bunch has a longitudinally flat-top profile with extent $-20\pi <k_p\zeta < 0$, and along each transverse axis a Gaussian profile with rms width of $k_p\sigma_r=3$ and a momentum spread of $\sigma_{pr}/(m_bc)=0.05$. The momentum spread is related to the normalised emittance $\epsilon_N$ (geometric emittance times the Lorentz factor) by $\sigma_{pr}/(m_b c) = \epsilon_N/\sigma_r$. The peak density of the bunch positrons and electrons is $n_b/2=0.02\,n_p$, i.e.\ $n_b$ is the total peak density of the bunch. The bunch propagation through a uniform plasma is considered in the co-moving frame $\zeta= z-u_b t,\,\tau= z/u_b$, with $\zeta$ the bunch slice, $u_b$ the bulk velocity of the bunch and $\tau$ the propagation time in plasma. The grid size is $k_p\Delta(x,y,\zeta)=(0.01,0.01,0.1)$ and the propagation step is $k_p\Delta z=2$. The bunch species and plasma electrons are represented by 16 and 4 macroparticles per cell, and the plasma ions are stationary. Adding more macroparticles per cell for the cold plasma species has no observable effect. Changing the number of macroparticles for the bunch does not affect the instability growth, although it decreases the initial wakefield.

From theory, the filamentation growth rate increases with transverse wavenumber for a cold bunch. In simulations, the finite spatial resolution limits the maximum wavenumber which can be modeled. This leads to a dominant wavenumber determined by the cell size. For the finite emittance considered in \cref{fig:tts_3d_intro}, diffusion results in a physical reduction of the growth rate at higher wavenumbers, yielding a dominant wavenumber well within the resolution limit of the simulation. In the next section, an analytical model is developed for wakefield-driven two-stream instabilities.

\section{\label{sec:theory_wf_cold}Filamentation of Cold Beams}

The charge density of the bunch $\rho_b$ drives an electrostatic plasma response, expressed as the longitudinal $E_z$ and transverse $\bm{W}_\perp$ wakefield \citep{Katsouleas1987}. A sinusoidal perturbation is assumed, $\rho_b=q_b \delta n_b g(x,y)$, where $\delta n_b$ is the amplitude of the density modulation and $g(x,y)=\tilde{g}(x,y)\cos{(k_{x} x+\varphi_x)}\cos{(k_{y} y+\varphi_y)}$ is the transverse profile, with $\tilde{g}$ a slowly-varying envelope, and $k_{x,y}$ and $\varphi_{x,y}$ the perturbation wavenumbers and phases along the transverse axes. The wakefield can be calculated from the wave equations, as detailed in \cref{sec:app_der_wake}.  Neglecting the bulk return current of the plasma electrons and assuming stationary plasma ions yields \citep{Keinings1987,Katsouleas1987}
\begin{nalign}\label{eq:wake_cos}
  E_z &= \frac{q_b\delta n_b}{\varepsilon_0} \frac{k_e^2 g(x,y)}{k_e^2+k_r^2}\int_\zeta^0 \diff \zeta' f(\zeta') \cos{k_e(\zeta-\zeta')} \\
  \bm{W}_\perp &= \frac{q_b \delta n_b}{\varepsilon_0}\frac{k_e \nabla_\perp g(x,y)}{k_e^2+k_r^2} \int_\zeta^0 \diff \zeta' f(\zeta')\sin{k_e(\zeta-\zeta')},
\end{nalign}
with $k_e=k_pc/u_b$ the electron wavenumber, $k_r=(k_x^2+k_y^2)^{1/2}$, and $f(\zeta)$ the longitudinal profile. The linear regime requires $\delta n_b\ll n_b$ and $|E_z|,|\bm{W}_\perp|\ll E_0$, with $E_0=m_e \omega_p c / e$ the non-relativistic wave-breaking field.

The local self-fields can be neglected for relativistic bunches, as the electric charge repulsion is compensated by the magnetic field due to the bunch current. The wakefield acts back on the bunch, where particles are accelerated or decelerated by $E_z$ and focussed or defocussed by $\bm{W}_\perp$.
The evolution of a cold bunch is described by the linearised fluid equation \citep{Chen2016}
\begin{equation}\label{eq:dn_obi_cold}
  \partial_\tau^2 \delta n_b =  \frac{2\omega_\beta^2}{q_b/\varepsilon_0}\left(\frac{\partial_z E_z}{\gamma_b^2} + \nabla_\perp \cdot \bm{W}_\perp \right).
\end{equation}

The evolution of the perturbation within the bunch can be found along its length as it propagates in plasma by a Laplace transform of the Green's function to \cref{eq:dn_obi_cold}  (\cref{sec:app_der_growth_cold}). For a longitudinal flat-top bunch with the head at $\zeta=0$, the growth of the modulation amplitude with respect to its initial value $\delta n_{b0}$ is
\begin{ngather}\label{eq:growth_obi}
  \Gamma_\mathrm{TS} = \frac{\delta n_{b,\mathrm{TS}}}{\delta n_{b0}} = \left|\sum_{n=0}^\infty \frac{\left[i\eta_u \tilde{g}(x,y) k_e|\zeta|\omega_\beta^2\tau^2\right]^n}{n!(2n)!}\right| \\
  \eta_u = \frac{(c^2-u_b^2)k_p^2 + u_b^2 k_r^2}{c^2 k_p^2+u_b^2 k_r^2}.
\end{ngather}
The first and second summand in the spectral factor $\eta_u$ represent the respective contribution of the longitudinal and transverse wakefield component and, therefore, of \gls{tsi} and \gls{tts}. The spectral dependency agrees with the analytical expression for \gls{obi} in \citep{Fainberg1969, Bret2010}. The series can be asymptotically expressed by $\delta n_{b,\mathrm{TS}}\approx [\delta n_{b0}/\sqrt{4\pi}] \exp{(\Gamma_{\infty})}/\sqrt{\Gamma_{\infty}}$, with $\Gamma_{\infty} = (3^{3/2}/2^{5/3})[\eta_u \tilde{g}(x,y) k_e|\zeta|\omega_\beta^2\tau^2]^{1/3}$. In the non-relativistic and ultra-relativistic limit for streams, the asymptotic form simplifies to previous works \citep{Jones1983, Claveria2022}. The phase velocity of the growing electrostatic wave reduces relative to the bunch velocity as (\cref{sec:app_der_growth_cold})
\begin{nalign}\label{eq:reduced_phase_vel}
  u_\psi&=u_b\left[1-\frac{2}{3^{3/2}}\frac{\Gamma_\infty}{\omega_p\tau}\right].
\end{nalign}

In addition to the filamentation instability, a single-species bunch is subject to the axisymmetric \gls{smi}, for which the spectral factor $\eta_u$ in \cref{eq:growth_obi} is substituted by Bessel functions \citep{Schroeder2011, Pukhov2011}. The growth rate of a transverse modulation within the bunch exceeds the rate at which the transverse envelope changes for $\sigma_r\gtrsim 3/k_r$. Although a quasineutral bunch is not subject to \gls{smi}, it requires the consideration of the filamentation instability for two species. When the mass of the bunch particles is equal, the introduced theory can be readily applied by defining $n_b$ as the total bunch density summed over all bunch species. For different particle masses, the fluid equations couple asymmetrically, with the filamentation of the two species developing at different rates \citep{Graw2022}. This would require an extension of the model presented here.

In order to test this analytic description of two-stream instabilities, comparisons are made to simulations. Two-dimensional simulations are used, in which relativistic beam particles effectively have one degree of freedom and $k_r=k_y$. The two-dimensional simulations are carried out with the electromagnetic \gls{pic} code OSIRIS \citep{Fonseca2002}. The grid size of the simulation is $k_p\Delta(y,z)=(0.02,0.06)$, and the time step is set to $\omega_p\Delta t=0.0172$. The bunch and plasma species are each represented by 384 and 192 macroparticles per cell. The number of particles per cell is significantly higher than that used in the three-dimensional simulations due to the relative decrease in the total number of cells in the two-dimensional simulations. The boundary conditions are open for the macroparticles and electromagnetic fields. The bunch is initialised in a vacuum and propagates into plasma. The bunch parameters are equivalent to \cref{fig:tts_3d_intro}, but with an initially cold beam. For an ideal cold bunch, there is no charge perturbation from which filamentation can develop. The bunch charge density is therefore transversely modulated with an amplitude of $e\delta n_b=0.01\sqrt{2}\,en_p$ and a wavenumber of $k_y/k_p=\pi$ to allow filaments to develop.

\begin{figure}
  \includegraphics[width=\linewidth]{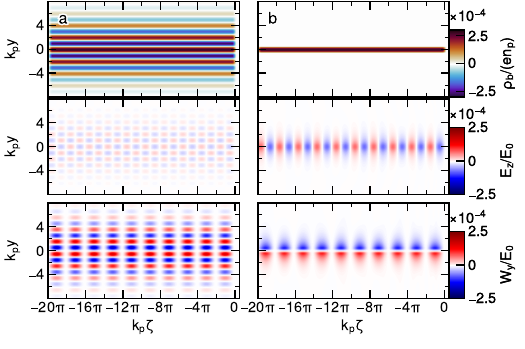}
  \caption{\label{fig:smi_tts_illustr} Bunch-driven wakefield immediately after entering the plasma, $\tau=0$. Bunch charge density and corresponding longitudinal and transverse wakefield for a a) wide quasi-neutral bunch with transverse modulation and b) narrow single-species bunch of identical amplitude $\delta n_b$ and width $k_y$.}
\end{figure}

\Cref{fig:smi_tts_illustr}a) shows the initial wakefield driven by the transversely modulated bunch when each bunch slice just entered the plasma, $\tau=0$. The longitudinal and transverse wakefield exhibit a longitudinal modulation at $k_\zeta=k_e$ and a transverse modulation at the seeded wavenumber $k_y=\pi k_p$. The transverse wakefield is stronger than the longitudinal component in agreement with the theoretical ratio $\tilde{\bm{W}}_\perp = \tilde{E}_z \bm{k}_r u_b/(k_p c)$ from \cref{eq:wake_cos}. For comparison, the wakefield driven by a narrow single-species bunch is shown in \cref{fig:smi_tts_illustr}b). Unlike the wide bunch, the wakefield extends beyond the narrow bunch. However, in both cases, the transverse wakefield periodically alternates between focussing and defocussing along the bunch, which gives rise to \gls{tts} for a transversely modulated bunch or \gls{smi} for a narrow single-species bunch.

\begin{figure}
  \includegraphics[width=\linewidth]{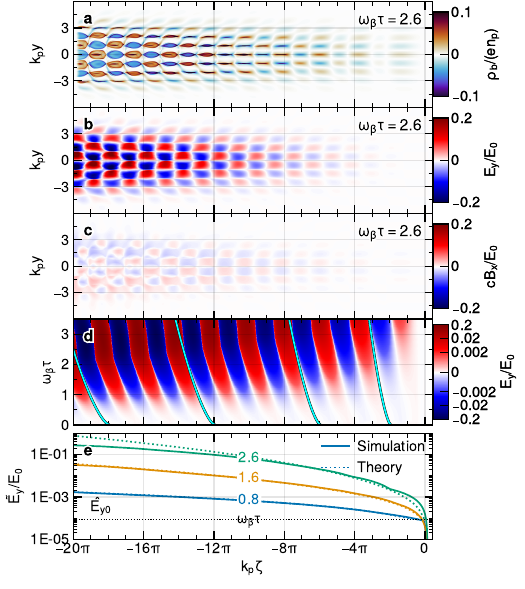}\\
  \includegraphics[width=0.52\linewidth]{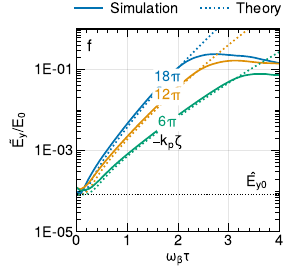}
  \includegraphics[width=0.46\linewidth]{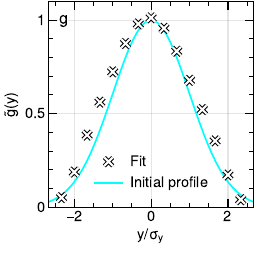}
  \caption{\label{fig:tts_azeta} The evolution of a modulated bunch propagating in plasma. The a) bunch charge density, b) electric and c) magnetic field of a filamented bunch with a transverse modulation, $k_y/k_p=\pi$, at $\omega_\beta\tau=2.6$. d) The transversely averaged electric field along the plasma length with the theoretical phase shift superimposed. e) The envelope growth of the electric field $\tilde{E}_y$, along the length of the bunch $\zeta$ at the propagation times $\omega_\beta\tau=\{0.8,1.6,2.6\}$, showing the simulation and theoretical values by solid and dotted lines, respectively. The black dotted line indicates the seed of the electric field. f) Simulated and theoretical envelope growth of the electric field along the propagation $\tau$ at equidistant bunch slices $k_p\zeta=\{-6\pi,-12\pi,-18\pi\}$. g) Bunch profile from simulation, obtained by fitting the observed growth to \cref{eq:growth_obi}, compared to the initial bunch shape.}
\end{figure}

The resulting growth of the filamentation instability from the initial plasma response in \cref{fig:smi_tts_illustr}a) is illustrated in \cref{fig:tts_azeta} at a propagation of $2.6/\omega_\beta$ in plasma. The modulation amplitude of the bunch charge density in \cref{fig:tts_azeta}a) increases along the bunch length, and contains a longitudinal modulation at $k_\zeta=k_e$ due to the electrostatic plasma response. The transverse wakefield from \cref{fig:tts_azeta}b) and c) alternates between focusing and defocusing, both transversely and along the bunch, resulting in alternating positron and electron filaments. The magnetic field in \cref{fig:tts_azeta}c) is weaker than the electric field by an order of magnitude and is predominantly due to the local bunch current. For a relativistic bunch, Coulomb repulsion is compensated by the magnetic field, so the beam evolution is determined entirely by the plasma wakefield.

The electric field (taken as the average over the range $0<k_y y<\pi$) in \cref{fig:tts_azeta}d) shows the growth along the bunch length as the bunch propagates in plasma. The modulation shifts backwards, illustrating that the phase velocity is lower than the bunch velocity. The superimposed lines represent the integral of the phase velocity from \cref{eq:reduced_phase_vel} over the length of the plasma and agree well with the phase of the wave.

\Cref{fig:tts_azeta}e) and f) show the envelope growth of the electric field (averaged over the range $-\pi<k_y y<\pi$) along the bunch and the plasma length, respectively. The seed value agrees well with the analytic expression for the Fourier spectrum $\hat{E}_{y0}=\mathcal{F}_\perp\{E_{y0}\}=[e\delta n_{b0}/\varepsilon_0] k_y/(k_p^2+k_y^2)$, obtained by solving \cref{eq:wake_cos} for the initial bunch profile. The growth of the electric field is compared with the semi-analytic solution to \cref{eq:growth_obi}, including the first ten terms, and shows excellent agreement along the bunch up to a propagation time in plasma of $\sim 2/\omega_\beta$.

To demonstrate the effect of a slowly varying transverse envelope on the growth, \cref{eq:growth_obi} is fitted to the simulation data along the plasma length at $k_p\zeta=-12\pi$ with $\tilde{g}(y)$ as a free parameter. The fit coefficient agrees well with the Gaussian profile of the bunch in \cref{fig:tts_azeta}g). In contrast to a longitudinal extent resulting in an increase of the growth along the bunch, the growth rate and seed level correlate with the transverse envelope $\tilde{g}(y)$. The growth rate at a given transverse coordinate can be treated as a stream with the local bunch density. The curved phase fronts in the bunch modulation are due to the dependency of the phase velocity on the transverse envelope, $\tilde{g}(y)^{1/3}$, in \cref{eq:reduced_phase_vel}.

Simulations show that beyond $\omega_\beta\tau = 2$, the field growth begins to decrease relative to the analytical predictions (\cref{fig:tts_azeta}d,f) while the phase velocity increases (\cref{fig:tts_azeta}e). This saturation is due to the beam becoming fully modulated, with the electron and positron filaments fully separating, as seen in \cref{fig:tts_azeta}a) for $k_p\zeta \leq -10\pi$.

\section{\label{sec:theory_wf_warm}Filamentation of Warm Beams}

The filamentation of the bunch depicted in \cref{fig:tts_3d_intro} results in a dominant wavenumber, a behaviour the theory for cold bunches cannot describe. Diffusion of warm bunches causes fine-scale perturbations within the bunch to spread out, reducing the growth rate of the instability. For non-relativistic temperatures, $\sigma_{pr}^2/(m_b c)^2\ll 1$, the fluid equation in \cref{eq:dn_obi_cold} can be modified to include the thermal pressure arising due to the transverse momentum spread (\cref{sec:app_der_growth_warm})
\begin{equation}\label{eq:dn_obi}
  \left(\partial_\tau^2+\frac{2}{3}\frac{\sigma_{pr}^2 k_r^2}{m_b^2\gamma_b^2}  \right) \delta n_b =  \frac{2\omega_\beta^2}{q_b/\varepsilon_0}\left(\frac{\partial_z E_z}{\gamma_b^2} + \nabla_\perp \cdot \bm{W}_\perp \right),
\end{equation}
where diffusion acts to damp transverse density modulations \citep{Bret2006}. Since all bunch slices are equally affected by diffusion, damping is purely temporal and can be treated separately from the spatiotemporal growth of the filamentation instability. In the absence of a wakefield, the exponential damping rate $\delta_D$ for a transverse perturbation in \cref{eq:dn_obi} is described by
\begin{nalign}
  \delta n_{b,D} &= \delta n_{b} \exp{(-\delta_D \tau)} \\
  \delta_D &= \sqrt{\frac{2}{3}}\frac{\sigma_{pr} k_r}{\gamma_b m_b}.
\end{nalign}
The total growth rate is, therefore, the sum of the growth rate from two-stream instabilities with the damping rate from diffusion, expressed by 
\begin{ngather}\label{eq:tot_growth}
    \Gamma_\mathrm{tot} = \delta n_b/\delta n_{b0} = \Gamma_\mathrm{TS}\exp{(-\delta_D \tau)}
\end{ngather}
The effect of temperature can only be considered as purely diffusive for $\sigma_{pr}/(m_b c)<[3/2^{10/3}(n_b/n_p)^{1/3}\gamma_b^{1/3}(1+\gamma_b^{-2})^{2/3}/(1+\gamma_b^{-1})^2]^{1/2}$ \citep{Bret2010b}. This corresponds to $\sigma_{pr}/(m_b c)<0.2$ for the bunch parameters in \cref{fig:tts_3d_intro}.

\begin{figure}
  \includegraphics{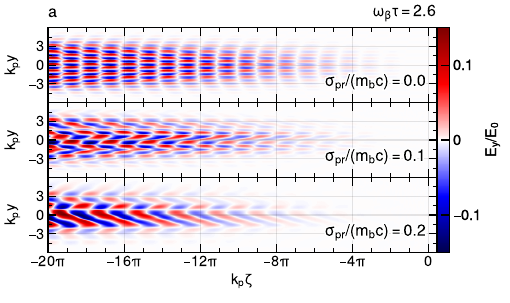}
  \includegraphics{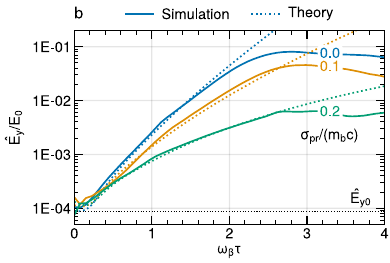}
  \caption{\label{fig:diff_atau}
  a) The transverse electric field resulting from a cold and warm bunches of increasing temperature with a transverse modulation at $k_y/k_p=2\pi$. b) The corresponding growth of the spectrum at the seeded wavenumber $\hat{E}_y=|\hat{E}_y|(k_{y})$, obtained from simulation and theory at different slices $\zeta$.
  }
\end{figure}

The influence of diffusion on the filamentation instability is examined for bunches with different temperatures. Since diffusion has a larger effect at higher wavenumbers, the parameters are as for the bunch in \cref{fig:tts_azeta} but with a transverse modulation at $k_y/k_p=2\pi$. The excited electric field is shown in \cref{fig:diff_atau}a) at $2.6/\omega_\beta$. The field is lower compared to \cref{fig:tts_azeta}b) due to the difference in wavenumber, agreeing with $\hat{E}_y \sim k_y/(k_p^2+k_y^2)$ from \cref{eq:wake_cos}. For the cold bunch, the seeded wavenumber continues to dominate along the length of the bunch.

For warm bunches, the phase fronts deviate from the curve given by the bunch profile. The field reduces with temperature close to the bunch head since the filamentation instability grows along the bunch while diffusion is spatially uniform. The transverse modulation shifts from the seeded wavenumber, a change that becomes evident further away from the bunch head. The growth of the field spectrum along the plasma length in \cref{fig:diff_atau}b) reveals that the seeded wavenumber is damped proportionally to the bunch temperature. The observation is in good agreement with the analytical description for the effect of diffusion on the growth in \cref{eq:tot_growth}.

\begin{figure}
    \centering
    \includegraphics[width=\linewidth]{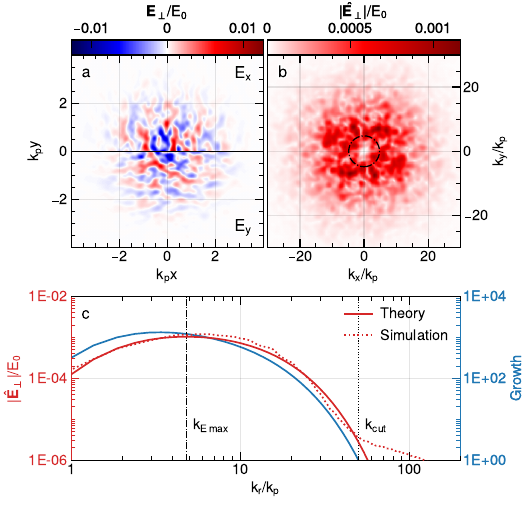}
    \caption{
    a) The transverse electric fields driven by the filamented bunch are shown as a transverse slice at $\omega_\beta\tau=2.6$, $k_p\zeta=-12\pi$. b) The corresponding 2D power spectrum, and c) the 1D power spectrum $|\hat{\bm{E}}_\perp|(k_r)$, showing the results from simulation and theory, as well as the theoretical growth $\Gamma$. The dotted and dash-dotted lines indicate the theoretical value for the wavenumber of maximum electric field amplitude $k_{E_\mathrm{max}}$, and the cut-off wavenumber $k_\mathrm{cut}$.
    }
    \label{fig:fft_3d}
\end{figure}

The development of filaments with wavenumbers lower than the seeded wavenumber indicates a higher growth rate for larger-scale filaments, such that the whole transverse spectrum of the instability has to be considered. In order to investigate the variation of the filamentation wavenumber, the electric fields corresponding to the transverse slice at $k_p\zeta=-12\pi$ in \cref{fig:tts_3d_intro} are shown in \cref{fig:fft_3d}a). The transverse component $E_y$ is predominantly modulated along $y$, and $E_x$ is predominantly modulated along $x$. However, transverse modulations occur with a broad range of spatial scales and orientations in the transverse plane.

For unseeded bunches, the instability grows from fluctuations in the bunch due to the finite temperature, and the resulting electric field is a superposition of all growing transverse modulations. The respective contributions of the wavenumbers can be separated by a Fourier transform. Taking the two-dimensional Fourier transform of the transverse electric field components and plotting the absolute amplitude, i.e.\ $|\hat{\bm{E}}_\perp|=|\hat{E}_y|+|\hat{E}_x|$, in \cref{fig:fft_3d}b) reveals a wide range of growing transverse wavenumbers. The spectrum is azimuthally symmetric, showing that growing transverse modulations have no preferred orientation in the transverse plane. The radial symmetry is in agreement with the spectral factor in \cref{eq:growth_obi}, $\eta_1\rightarrow k_r^2/(k_p^2+k_r^2)$, which predicts that the growth rate of the filamentation instability only depends on the absolute value of the transverse wavevector. Thus, the filamentation in transverse planes is coupled, and the transverse modulations in each plane cannot be treated independently.

Averaging the spectrum of the electric field in \cref{fig:fft_3d}b) over all orientations, $(k_x,k_y)\rightarrow k_r$, gives the radial spectrum in \cref{fig:fft_3d}c). The spectrum of the electric field grows with transverse wavenumbers up to $k_r/k_p \sim 5$ due to the higher growth rate of the filamentation instability and reduces for higher wavenumbers due to diffusion. The comparison to theory requires an analytical description of the fields at $\tau=0$, which act to seed the instability. These seed fields are found from simulation to scale as $|\hat{\bm{E}}_{\perp 0}| \sim (k_r \sigma_{pr}^3)^{1/2}$, with the absolute value determined by the simulation. The wakefield after propagation is the product of the the seed spectrum with the theoretical growth spectrum, $|\hat{\bm{E}}_\perp|=|\hat{\bm{E}}_{\perp 0}(k_r)|\Gamma_\mathrm{tot}(k_r)$, which shows an excellent agreement to the simulation.

The spectrum of the growth exhibits a transverse wavenumber of maximum growth $k_{\Gamma_\mathrm{max}}(\tau,\zeta)$ and cut-off wavenumber $k_\mathrm{cut}(\tau,\zeta)$ above which the instability is suppressed. In the relativistic limit, the wavenumbers are numerically obtained by solving the following expressions for $k_{\Gamma_\mathrm{max}}$ or $k_\mathrm{cut}$ (\cref{sec:app_der_growth_warm})
\begin{nalign}\label{eq:krmax}
  2 \Gamma_\infty(k_{\Gamma_\mathrm{max}}) &= 3(1+k_{\Gamma_\mathrm{max}}^2)\delta_D(k_{\Gamma_\mathrm{max}})\tau +1 \\
  \Gamma_\infty(k_\mathrm{cut}) &= \delta_D(k_\mathrm{cut})\tau+ \ln{\sqrt{4\pi \Gamma_\infty(k_\mathrm{cut})}}.
\end{nalign}
The wavenumber of maximum growth scales by $k_{\Gamma_\mathrm{max}}\sim \sigma_{pr}^{-1/3}$ and the cut-off wavenumber scales by $k_\mathrm{cut}\sim \sigma_{pr}^{-1}$ with the bunch temperature. Since the two-stream instability is spatiotemporal, while diffusion is spatially uniform, the characteristic wavenumbers depend on the propagation time in plasma and position within the bunch. For the scaling of the seed field, the wavenumber of maximum spectral value $k_{E_\mathrm{max}}(\tau,\zeta)$ is obtained from
\begin{equation}\label{eq:kEmax}
    1+3k_{E_\mathrm{max}}^2 + 4\Gamma_\infty(k_{E_\mathrm{max}}) = 6(1+k_{E_\mathrm{max}}^2)\delta_D(k_{E_\mathrm{max}}) \tau.
\end{equation}
The predicted wavenumber at which the electric field is maximum, $k_{E\max}\approx 4.9$, from \cref{eq:kEmax} aligns well with the simulation data. The electric field above the calculated cut-off wavenumber, $k_r/k_p\gtrsim 50$, is attributed to numerical noise.

\begin{figure}
  \includegraphics{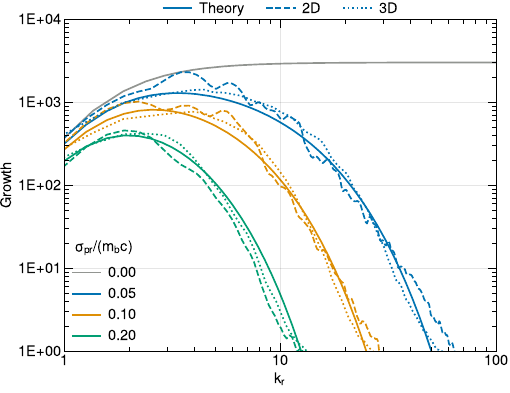}
  \caption{\label{fig:tts_gr_temp}
  The spectral growth dependency on the bunch temperature. The theoretical growth for different bunch temperatures at $\omega_\beta\tau=2.6$ and $k_p \zeta=-12\pi$ (solid lines), compared to 2D electromagnetic (dashed lines) and 3D quasi-static (dotted lines) PIC simulations.
  }
\end{figure}

The whole scope of the introduced theory is compared to two- and three-dimensional simulations of unseeded warm bunches with different temperatures in \cref{fig:tts_gr_temp}. Other parameters are as for the bunch in \cref{fig:tts_3d_intro}. The growth spectrum from simulations is obtained by taking the ratio between the electric field spectrum after propagation with the scaling of the seed field. This ratio is aligned to the growth spectrum from theory for two- and three-dimensional simulations, respectively. Small variations in the field spectrum can occur when the filamentation instability grows from random fluctuations in the bunch. Thus, the growth spectrum is averaged over five two-dimensional runs and three three-dimensional runs for each temperature and compared to the analytical expression for the total growth in \cref{eq:tot_growth}.

Agreement is found for the dependency of the growth spectrum on the temperature for both two- and three-dimensional simulations. The alignment is better in three dimensions since the total number of bunch particles is an order of magnitude higher. For cold bunches, theory predicts that the growth increases with wavenumber due to the filamentation instability. For warm bunches, the growth increases with wavenumber up to $k_{\Gamma_\mathrm{max}}$ and then decreases as the influence of diffusion becomes stronger. With higher temperatures, the growth is lower for all wavenumbers and the wavenumber of maximum growth and cut-off wavenumber shift to lower values in good agreement with the predicted values from evaluating \cref{eq:krmax}. Thus, transverse modulations in the bunch occur at larger scales. The distance between filaments is inversely related to $k_{E_\mathrm{max}}$. However, this means that the in-plane distance is higher in three-dimensional simulations with $k_x\sim k_y\sim k_{E_\mathrm{max}}/\sqrt{2}$, compared to the distance in two-dimensional simulations with $k_y \sim k_{E_\mathrm{max}}$. 

The analytical expression accurately predicts the dependency of the growth from the wakefield-driven filamentation instability and the damping from diffusion. The theory also verifies that the growth of the filamentation instability can be effectively modelled in two dimensions at a lower in-plane wavenumber without losing generality.

\begin{figure}
  \includegraphics{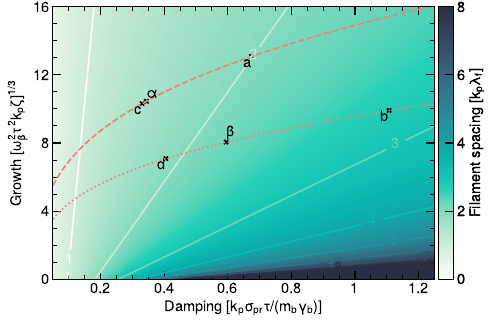}
  \caption{\label{fig:kEmax_exp}The distance between filaments as a function of bunch and plasma parameters. The y and x-axis are proportional to the growth of \gls{tts} and damping due to diffusion. The dashed line corresponds to the experimental parameters in \citep{Verra2024} for varying plasma density, while the dotted line corresponds to the experimental parameters in \citep{Allen2012}. Points a, b, c, and d correspond to individual measurements in \citep{Verra2024}, \citep{Allen2012}, with $\alpha$, $\beta$ marking the point at which the distance between filaments is predicted to reach the rms bunch width.}
\end{figure}

The expected distance between filaments, $\lambda_f = 2\pi/k_{E_\mathrm{max}}$, is shown in \cref{fig:kEmax_exp} as a function of the spatiotemporal growth and damping from diffusion.
At the back of the bunch, where filamentation is strongest, the expected distance between filaments is independent of the bunch length, depending instead on the total bunch charge, $\omega_\beta^2\zeta\sim \int n_b \diff \zeta$. While the theory developed in this work considers a longitudinally flat-top beam, this general dependence can readily be applied to bunches with arbitrary longitudinal profiles.

In experiments carried out with both proton \citep{Verra2024} and electron \citep{Allen2012} bunches, the onset of filamentation was studied by varying the plasma density. Taking the proton bunch parameters \footnote{$400\,\mathrm{GeV}$ proton bunch with a total charge of $43\,\mathrm{nC}$, an rms width $\sigma_r = 0.5\,\mathrm{mm}$, a normalised emittance of $2.5\,\mathrm{mm}\,\mathrm{mrad}$, and a longitudinally Gaussian profile with $\sigma_\zeta /c = 163\,\mathrm{ps}$ ($\int\omega_\beta^2\diff\zeta=(2\pi)^{1/2}\omega_\beta^2\sigma_\zeta$). The plasma length was $c\tau=10\,\mathrm{m}$} in \citep{Verra2024} and varying the plasma density gives the dashed line in \cref{fig:kEmax_exp}. Point (a) corresponds to a plasma density $n_p=9.38\E{14}\,\mathrm{cm}^{-3}$, for which filamentation was observed. The predicted distance between filaments, $\lambda_f=2/k_p=0.34\,\mathrm{mm}$, is comparable to the observed distance of $0.27\,\mathrm{mm}$. Taking the electron bunch parameters \footnote{$0.06\,\mathrm{GeV}$ electron bunch with a total charge of $1\,\mathrm{nC}$, an rms width $\sigma_r = 0.065\,\mathrm{mm}$, a normalised emittance of $6\,\mathrm{mm}\,\mathrm{mrad}$, and an rms length $\sigma_\zeta /c = 5\,\mathrm{ps}$. The plasma length was $c\tau=0.02\,\mathrm{m}$} in \citep{Allen2012} and varying the plasma density gives the dotted line in \cref{fig:kEmax_exp}. Point (b) corresponds to a plasma density $n_p=12\E{16}\,\mathrm{cm}^{-3}$, for which filamentation was observed. The predicted distance between filaments, $\lambda_f=2.7/k_p=0.042\,\mathrm{mm}$, appears to give agreement with the observed filamentation distance, although the transverse bunch envelope observed in the experiment was significantly modified through its interaction with the plasma and no longer resembles a Gaussian ellipsoid.

The points (c) and (d) correspond to the cases in \citep{Verra2024}, \citep{Allen2012} where a low plasma density was used, and filamentation was suppressed. For (c), approximately $50\,$\% of shots led to filamentation, suggesting that this is the threshold for the instability. For (d), no filamentation was observed. This threshold for filamentation correlates with the
predicted distance between filaments exceeding the rms width of the bunch. The points ($\alpha$) and ($\beta$) correspond to the cases in \citep{Verra2024}, \citep{Allen2012}, where the predicted distance between filaments is equal to the rms bunch width. Point (c), with a plasma density of $n_p=2.25\E{14}\,\mathrm{cm}^{-3}$, is close to ($\alpha$), with a plasma density of $2.44\E{14}\,\mathrm{cm}^{-3}$. Point (d), with a plasma density of $n_p=1.6\E{16}\,\mathrm{cm}^{-3}$ is well below point ($\beta$), with a plasma density of $n_p=3.4\E{16}\,\mathrm{cm}^{-3}$. The distance between filaments for the instability cutoff, $2\pi/k_\mathrm{cut}$, corresponds to a plasma density 50--140 times lower than the observed threshold. This dependence of the instability threshold on $k_{E_\mathrm{max}}$ and not $k_\mathrm{cut}$ may be due to competition of the filamentation instability with \gls{smi} of the charged bunches used in these experiments. Further experimental and numerical studies would allow this prediction for the instability threshold to be tested across a larger parameter space.

\section{\label{sec:concl}Conclusion}

A three-dimensional, spatiotemporal theory for the wakefield-driven filamentation instability is presented for warm bunches of finite size. The weakly and strongly relativistic regimes, referred to as \gls{tsi} and \gls{tts}, arise from the longitudinal and transverse wakefield components. In the limit of a cold stream, the analytical expressions for \gls{tsi} and \gls{tts} simplify to previous works. The electrostatic plasma response leads to the growth of transverse filaments with an additional longitudinal modulation. The transverse bunch profile influences both the growth rate and the seed level, with the growth rate at a fixed transverse position being equivalent to a stream with the local bunch density.

For beams with finite emittance, diffusion acts to damp small-scale filamentation. The dependency of the growth spectrum on the temperature is identified for dilute bunches. Theory and simulations show that the filamentation growth rate depends on $k_r=(k_x^2+k_y^2)^{1/2}$. Two-dimensional simulations reproduce the behaviour of three-dimensional simulations in the linear regime, with the caveat that $k_r=k_y$ in this reduced geometry, resulting in filaments that are more tightly clustered. Explicit expressions for the dominant and cut-off wavenumber are calculated and depend on the propagation time in plasma and position within the bunch. This arises as diffusion is spatially uniform while the filamentation instability grows along the bunch length. Remarkable agreement is found between theory and \gls{pic} simulations.

Although the analytical treatment developed here considers a longitudinally flat-top beam, a general dependence on the expected distance between filaments is found for bunches with arbitrary profile. The predicted distance between filaments gives good agreement with previously published experimental results. For single-species beams, filamentation appears to be suppressed when the predicted distance between filaments is larger than the rms beam width. These findings provide a crucial basis for designing laboratory astrophysics experiments investigating filamentation instabilities and for PWFA experiments seeking to avoid them.

\begin{acknowledgments}

We thank Patric Muggli for helpful discussions related to this work.

Funded by the Deutsche Forschungsgemeinschaft (DFG, German Research Foundation) under Germany´s Excellence Strategy – EXC 2094 – 390783311.
\end{acknowledgments}

\appendix

\section{Derivation of Wakefield-Driven Two-Stream Growth for Warm Bunches}\label{sec:app_der_ts}

\subsection{Wakefield Induced by a Modulated Bunch}\label{sec:app_der_wake}

A dilute bunch propagating in the $+z$ direction through an unmagnetised plasma leads to an electrostatic plasma response \citep{Katsouleas1987, Keinings1987, Lawson1977, Bret2004}. The associated fields are $E_z$, $\bm{W}_\perp=\bm{E}_\perp + u_b \bm{\hat{z}}\times\bm{B}_\perp$, with $\bm{\hat{z}}$ the unit vector along z and $u_b$ the bunch velocity. Only the oscillatory plasma current $\bm{j}_p$ is considered, and the small bulk return current for underdense bunches is neglected. Therefore, Ohm's law reduces to $\mu_0\partial_t \bm{j}_p=-k_p^2\bm{E}$, with $\mu_0$ the vacuum permeability and $k_p$ the plasma wavenumber. The fields can then be described by \citep{Keinings1987}
\begin{nalign}
  \left(\nabla^2-\partial_t^2/c^2-k_p^2\right)\bm{E} &= \mu_0\partial_t \bm{j}_b + \nabla (\rho_b+\delta \rho_p)/\varepsilon_0\\
  \left(\nabla^2-\partial_t^2/c^2-k_p^2\right)\bm{B} &= -\mu_0\nabla \times \bm{j}_b,
\end{nalign}
with $\rho_b$ the bunch charge density, $\delta \rho_p$ the charge density of the plasma perturbation, $\bm{j}_b= \rho_b u_b \bm{\hat{z}}$ the bunch current density, $c$ the speed of light and $\varepsilon_0$ the vacuum permittivity. The plasma perturbation connects to the bunch charge density via its fluid equation, $m_e\partial_t^2 \delta\rho_p=-e^2 \nabla\cdot\bm{E}=-e^2(\rho_b+\delta \rho_p)/\varepsilon_0$, where $m_e$ is the electron mass.

With bunch slice $\zeta$ and propagation time in plasma $\tau$, the Lagrangian frame of the bunch is defined by $\zeta= z-u_b t,\,\tau= z/u_b$. The partial derivatives transform in the bunch frame to $\partial_t=-u_b \partial_\zeta$ and $\partial_z=\partial_\zeta+\partial_\tau/u_b$. Assuming that the bunch evolution along its propagation $z$ is significantly slower than the response of the plasma electrons along the bunch $\zeta$, the quasi-static approximation for the plasma quantities and wakefield can be assumed. Therefore, $|\partial_\zeta \delta \rho_p|\gg |\partial_\tau \delta \rho_p/u_b|$ and $(\partial_z^2-\partial_t^2/c^2)\delta \rho_p\to (1-u_b^2/c^2)\partial_\zeta\delta \rho_p=\partial_\zeta^2\delta\rho_p/\gamma_b^2$, with $\gamma_b=(1-u_b^2)^{-1/2}$ the Lorentz factor. The same applies for $\bm{E}$ and $\bm{B}$.

Defining the 3D Fourier transform
\begin{nalign}
    \hat{\rho}_b&=\mathcal{F}_{\zeta xy}\{\rho_b\}(k_\zeta,k_x,k_y)\\&=\iiint_{-\infty}^\infty \diff\zeta\diff x\diff y \rho_b \exp\left(-\im k_\zeta\zeta-\im k_x x-\im k_y y\right),
\end{nalign}

the spectral form of the plasma fluid in the bunch frame is given by $\delta \hat{\rho}_p=-k_e^2 \hat{\rho}_b/(k_\zeta^2+k_e^2)$, with $k_e=ck_p/u_b$. The field components transform to
\begin{nalign}\label{eq:app_spectr_eb}
  \hat{E}_z &= -\frac{\im}{\varepsilon_0\sqrt{2\pi}} \frac{k_\zeta(k_\zeta^2/\gamma_b^2 + k_p^2)\mathcal{F}_{\zeta xy}\{\rho_b\}}{(k_\zeta^2-k_e^2)(k_\zeta^2/\gamma_b^2+k_p^2+k_r^2)} \\
  \bm{\hat{E}}_\perp &= \frac{1}{\varepsilon_0\sqrt{2\pi}}\frac{k_\zeta^2\mathcal{F}_{\zeta xy}\{\nabla_\perp\rho_b\}}{(k_\zeta^2-k_e^2)(k_\zeta^2/\gamma_b^2+k_p^2+k_r^2)} \\
  \bm{\hat{B}}_\perp &= \frac{u_b/c^2}{\varepsilon_0\sqrt{2\pi}} \frac{\mathcal{F}_{\zeta xy}\{\nabla^\perp \rho_b\}}{k_\zeta^2/\gamma_b^2+k_p^2+k_r^2}.
\end{nalign}
with $k_r=(k_x^2+k_y^2)^{1/2}$, $\nabla_\perp=(\partial_x,\partial_y)$ and $\nabla^\perp=(-\partial_y,\partial_x)$.

To obtain the fields for a small-scale perturbation in 3D configuration space, a quasi-neutral bunch (equal populations of particles with opposite charge) is superimposed by a non-neutral transverse modulation
$g(x,y)=\tilde{g}(x,y)\cos{(k_xx+\varphi_x)}\cos{(k_yy+\varphi_y)}$, with $\tilde{g}$ the slowly varying transverse envelope, i.e.\ $|\partial_y \tilde{g}|\ll k_y |\tilde{g}|$, and $k_{x,y}$ and $\varphi_{x,y}$ the respective modulation wavenumbers and phases. The positron density may be given by $n_{bp}= [n_b/2]\tilde{f}(\zeta)\tilde{g}(x,y)+[\delta n_b/\sqrt{2}] f(\zeta)\tilde{g}(x,y)\cos{(k_x x)}\cos{(k_y y-\pi/4)}$ and respectively for the electron density $n_{be}=[n_b/2]\tilde{f}(\zeta)\tilde{g}(x,y)+[\delta n_b/\sqrt{2}] f(\zeta)\tilde{g}(x,y)\cos{(k_x x)}\sin{(k_y y-\pi/4)}$, with $n_b$ the total density amplitude of the bunch and $\delta n_b$ the amplitude of the density perturbation. The longitudinal bunch shape and its slowly varying envelope are given by $f(\zeta)$ and $\tilde{f}(\zeta)$. The net charge density of the bunch, $\rho_b=q_b\delta n_b f(\zeta)g(x,y)$, with $q_b$ the charge of the bunch particles, serves as the source for the fields. The inverse Fourier transforms for $\zeta<0$ are
\begin{widetext}
\begin{nalign}
  \hat{E}_z &= \frac{q_b\delta n_b}{\varepsilon_0} \frac{\mathcal{F}_{xy}\{g(x,y)\}}{k_e^2+k_r^2} \int_{-\infty}^0\diff \zeta' f(\zeta')\left[ k_e^2\cos{k_e(\zeta-\zeta')} + k_r^2\exp\left(-\gamma_b\sqrt{k_p^2+k_r^2}|\zeta-\zeta'|\right) \right] \\
  \bm{\hat{E}}_\perp &= \frac{q_b\delta n_b}{\varepsilon_0}\frac{\mathcal{F}_{xy}\{\nabla_\perp g(x,y)\}}{k_e^2+k_r^2} \int_{-\infty}^0\diff \zeta' f(\zeta')\left[ k_e\sin{k_e(\zeta-\zeta')}-\gamma_b\sqrt{k_p^2+k_r^2}\exp\left(-\gamma_b\sqrt{k_p^2+k_r^2}|\zeta-\zeta'|\right) \right] \\
  \bm{\hat{B}}_\perp &= \frac{\gamma_b u_b}{c^2}\frac{q_b \delta n_b}{\varepsilon_0}\frac{\mathcal{F}_{xy}\{\nabla^\perp g(x,y)\}}{\sqrt{k_p^2+k_r^2}}\int_{-\infty}^0\diff \zeta'  f(\zeta')\exp\left(-\gamma_b\sqrt{k_p^2+k_r^2}|\zeta-\zeta'|\right).
  \end{nalign}
\end{widetext}
Neglecting the small spectral broadening due to $\tilde{g}(x,y)$, the transverse inverse Fourier transform for the transverse component of the electric field gives
\begin{equation}
  \bm{E}_\perp \sim\mathcal{F}_{xy}^{-1}\left\{\frac{\mathcal{F}_{xy}\left\{\nabla_\perp g(x,y)\right\}}{k_e^2+k_r^2}\right\} \approx \frac{\nabla_\perp g(x,y)}{k_e^2+k_r^2}
\end{equation}
and for the $z$ component gives $E_z\sim -k_e/(k_e^2+k_r^2)g(x,y)$.

The second electromagnetic summand in the integral can be split into the contribution of the local bunch slice and the inductive, purely decaying fields due to a change in bunch shape. The latter can be safely ignored if the plasma is non-diffusive \citep{Keinings1987}. The electromagnetic terms simplify to
\begin{equation}
    \int_\zeta^0\diff \zeta' f(\zeta')\exp\left( -\gamma_b\sqrt{k_p^2+k_r^2}|\zeta-\zeta'| \right) \approx \frac{f(\zeta)}{\gamma_b\sqrt{k_p^2+k_r^2}}.
\end{equation}

Without any limitation on the longitudinal shape, the fields can be expressed by
\begin{nalign}\label{eq:app_wake_cos}
  E_z &= \frac{q_b\delta n_b}{\varepsilon_0} \frac{k_e g(x,y)}{k_e^2+k_r^2} \int_\zeta^0 \diff \zeta' f(\zeta') k_e\cos{k_e(\zeta-\zeta')} \\
  \bm{E}_\perp &= \frac{q_b\delta n_b}{\varepsilon_0}\frac{\nabla_\perp g(x,y)}{k_e^2+k_r^2} \\ &\times \left[ \int_\zeta^0 \diff \zeta' f(\zeta')k_e\sin{k_e(\zeta-\zeta')} - f(\zeta) \right] \\
  \bm{B}_\perp &= -\frac{u_b}{c^2}\frac{q_b\delta n_b}{\varepsilon_0}\frac{\nabla^\perp g(x,y)}{k_p^2+k_r^2} f(\zeta).
\end{nalign}
For relativistic bunches $k_e\approx k_p$, the latter charge-repulsion term in $\bm{E}_\perp$ and the magnetic field $\bm{B}_\perp$ approximate to the local bunch contribution $W_f\sim f(\zeta)(1-u_b^2)=f(\zeta)/\gamma_b^2$ and are usually neglected for $\gamma_b\gg 1$. Each bunch slice $f(\zeta)$ drives a wakefield with an amplitude proportional to the transverse bunch shape. These contributions sum up along the bunch.

\subsection{Growth of Two-Stream Filamentation}\label{sec:app_der_growth_cold}

The excited fields act on the bunch. Assuming a cold bunch with a longitudinal momentum much larger than its transverse momentum, the linearised fluid equation gives
\begin{equation}\label{eq:app_fluid_nb_cold}
  \partial_\tau^2 \delta n_b=\left(\partial_t+u_b\partial_z\right)^2 \delta n_b = \frac{2\omega_\beta^2}{q_b/\varepsilon_0}
  \left(\frac{\partial_z E_z}{\gamma_b^2} + \nabla_\perp \cdot \bm{W}_\perp \right),
\end{equation}
with $\omega_\beta=[q_b^2 n_b/(2\gamma_b \varepsilon_0 m_b)]^{1/2}$ the betatron frequency and $m_b$ the mass of bunch particles.

The fields result in positive feedback, which gives rise to spatiotemporal growth. The growth due to the wakefield, given by the integral terms in $E_z$ and $\bm{E}_\perp$, can be tracked by applying the spatial derivative along $\zeta$ to \cref{eq:app_fluid_nb_cold}. In the strongly coupled regime, $\omega_\beta\tau\ll k_e\zeta$, the bunch perturbation can be described by $\delta n_b g(x,y)\tilde{f}(\zeta)[\exp{(\im k_e \zeta)}/2+\mathrm{c.c.}]$, considering the longitudinal wavenumber of the wakefields at $k_\zeta=k_e$. The integral along $\zeta$ from \cref{eq:app_wake_cos} reduces to
\begin{equation}
    \partial_\zeta \int_\zeta^0 \tilde{f}(\zeta')\frac{\exp{(\im k_e \zeta')}}{2}\sin{k_e(\zeta-\zeta')}\approx \frac{\im}{2}\tilde{f}(\zeta)\exp{(\im k_e\zeta)}.
\end{equation}
For a flat-top bunch, $\tilde{f}(\zeta)=\Theta(-\zeta)$, the initial perturbation is given by $\delta n_b(\tau=0,\zeta)=\delta n_{b0}\Theta(-\zeta)$. The local terms in \cref{eq:app_wake_cos}, which only act within a bunch slice $E_{x,y} \sim f(\zeta)$ and $B_{x,y} \sim u_b f(\zeta)$, are negligible compared to the growing wakefield term. For a slowly varying transverse envelope, the transverse gradient simplifies to $\nabla_\perp^2 g(x,y)\approx -k_r^2 g(x,y)$ and the perturbation amplitude follows \citep{Schroeder2011}
\begin{ngather}\label{eq:app_nb_green}
  \left[ \partial_\zeta \partial_\tau^2 + \im\eta_u k_e\omega_\beta^2\tilde{g}(x,y) \right] \delta n_b(\tau,\zeta) = 0 \\
  \eta_u = \frac{(c^2-u_b^2)k_p^2 + u_b^2 k_{r}^2}{c^2 k_p^2+u_b^2 k_{r}^2}.
\end{ngather}
The spectral parameter $\eta_u$ includes the dependency on the bunch velocity, representing the relative contribution of the (longitudinal) two-stream and transverse two-stream instability. The Green's function can be solved by a double Laplace transform \citep{Schroeder2011}
\begin{nalign}\label{eq:app_lapl_dn_obi}
    &\mathcal{L_{\zeta\tau}}\{\delta n_b\}(k_\zeta,k_\tau)=\iint_{-\infty}^\infty \diff \tau \diff \zeta \delta n_b \exp\left(-k_\zeta\zeta-k_\tau\tau\right) \\
    &=\frac{k_\zeta \mathcal{L}_\zeta\left\{(k_\tau +\partial_\tau)\delta n_b(\tau=0,\zeta)\right\}+\mathcal{L}_\tau\left\{\partial_\tau^2 \delta n_b(\tau,\zeta=0)\right\}}{k_\zeta k_\tau^2+\im \eta_u k_e\omega_\beta^2 \tilde{g}(x,y)}.
\end{nalign}
Assuming a sharp plasma boundary at $\tau=0$ sets the initial condition $\delta n_b(\tau,\zeta=0)=\delta n_{b0}$, which results in $\partial_\tau \delta n_b(\tau=0,\zeta)=\partial_\tau^2 \delta n_b(\tau,\zeta=0)=0$ and $\mathcal{L}_\zeta\{\delta n_b(\tau=0,\zeta)\}=\delta n_{b0}/k_\zeta$. Using the Residue theorem for the inverse Laplace transform in $\zeta$ and the relation $\mathcal{L}_\tau^{-1}\{\tau^{-2n-1}\}=t^{2n}/(2n)!$ \citep{Abramowitz1964} for the inverse transform in $\tau$ gives the solution to \cref{eq:app_lapl_dn_obi} as a complex power series for $\tau\geq0,\zeta\leq 0$
\begin{equation}\label{eq:app_ts_growth_exact}
  \delta n_{b,\mathrm{TS}} = \delta n_{b0}\sum_{n=0}^\infty \frac{\left[i\eta_u \tilde{g}(x,y) k_e|\zeta|\omega_\beta^2\tau^2\right]^n}{n!(2n)!}.
\end{equation}
The solution contains a growing imaginary and oscillatory real term, which can be obtained by taking the absolute value $\Gamma_\mathrm{TS}=|\delta n_b/\delta n_{b0}|$ and the phase $\psi(\delta n_b)$.
The asymptotic expansion, $\tau\rightarrow \infty$, to \cref{eq:app_ts_growth_exact} gives
\begin{align}
  \delta n_{b,\mathrm{TS}} &\approx \frac{\delta n_{b0}}{\sqrt{4\pi}} \frac{\exp \{(3/2^{2/3})[\im\eta_u \tilde{g}(x,y) k_e|\zeta| \omega_\beta^2\tau^2 ]^{1/3}\}}{\sqrt{(3/2^{2/3})[i\eta_u\tilde{g}(x,y) k_e|\zeta|\omega_\beta^2\tau^2 ]^{1/3}}}.
\end{align}
The growth of the bunch perturbation due to the combined two-stream instabilities is
\begin{nalign}\label{eq:app_ts_growth_asympt}
  \delta n_{b,\mathrm{TS}}&\approx \frac{\delta n_{b0}}{\sqrt{4\pi}} \frac{\exp{\Gamma_\infty}}{\sqrt{\Gamma_\infty}} \\
  \Gamma_\infty &= \frac{3^{3/2}}{2^{5/3}}\left[\eta_u \tilde{g}(x,y) k_e|\zeta|\omega_\beta^2\tau^2\right]^{1/3}.
\end{nalign}
The oscillatory term yields a phase $\psi$ and corresponding phase velocity $u_\psi= -\partial_t \psi/\partial_z \psi$ of the growing wave
\begin{nalign}
  \psi&=\frac{\pi}{4}-k_e |\zeta| - \frac{3}{2^{5/3}} (\eta_u  \tilde{g}(x,y) k_e|\zeta| \omega_\beta^2\tau^2)^{1/3} \\
  u_\psi&=u_b\left[1-\frac{1}{2^{2/3}}\left(\eta_u \tilde{g}(x,y) \frac{\omega_\beta^2}{\omega_p^2} \frac{|\zeta|}{c\tau} \right)^{1/3}\right],
\end{nalign}
comparable to the phase velocity from \gls{smi} \citep{Schroeder2011,Pukhov2011}.

\begin{figure}
  \includegraphics[width=\linewidth]{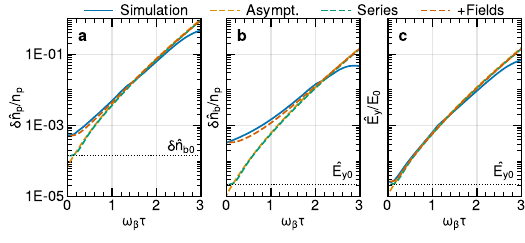}
  \caption{\label{fig:app_pne_pert} The growth of the spectral a) plasma perturbation, b) bunch perturbation and c) electric field at $k_p \zeta=-12\pi$ from simulation and theory for the cold bunch in \cref{fig:diff_atau}. The yellow and green dashed lines show the asymptotic (\ref{eq:app_ts_growth_asympt}) and semi-analytic (\ref{eq:app_ts_growth_exact}) solution, and the orange line includes the growth from the initial fields. The dotted lines indicate the seed for the two-stream instability.}
\end{figure}

At early propagation times, the initial bunch perturbation and, consequently, the plasma perturbation is much larger than the exponential growth of the two-stream instability. For short times, the growth evolves as
\begin{equation}\label{eq:app_ts_growth_seed}
    \delta n_{b,S} = \delta n_{b0}\left[\eta_u \tilde{g}(x,y)\omega_\beta^2\tau^2+1\right].
\end{equation}
This can be observed in \cref{fig:app_pne_pert}, where the bunch and plasma perturbation are not purely exponential. The same initial field dominates their initial growth, and the exponentially growing term only dominates after the bunch has propagated for some time. However, the transverse electric field, being the difference between plasma and bunch charge density
\begin{equation}
    \frac{\bm{E}_\perp}{E_0}=\frac{\bm{k}_r}{k_e^2+k_r^2}\frac{e \delta n_p-q_b\delta n_b}{n_p}
\end{equation}
exhibits exponential growth even at early times.

\subsection{Influence of Diffusion}\label{sec:app_der_growth_warm}

Extending \cref{eq:app_fluid_nb_cold} for warm bunches with a thermal spread $\sigma_{pr}$ requires the pressure term $\mathcal{P}$ to be included in \cref{eq:app_fluid_nb_cold}
\begin{nalign}
  \partial_\tau^2 \delta n_b = \frac{2\omega_\beta}{q_b/\varepsilon_0} \left(\frac{\partial_z E_z}{\gamma_b^2} + \nabla_\perp \cdot \bm{W}_\perp \right) +\frac{\nabla^2\mathcal{P}}{\gamma_bm_b},
\end{nalign}
where the pressure can be described by $\mathcal{P}=(2/3)\sigma_{pr}^2\delta n_b/(\gamma_b m_b)$ for non-relativistic temperatures, $\sigma_{pr}^2/(m_bc)^2\ll1$ \citep{Bret2006}. The thermal spread can be related to the normalised emittance $\epsilon_N$ by $\sigma_{pr}/(m_b c)=\epsilon_N/\sigma_r$, where $\sigma_r$ is the rms width for a Gaussian bunch. The effect of emittance-related diffusion is purely temporal. It can be considered separately from the wakefield-driven two-stream instability since all bunch slices are equally affected by the bunch divergence, and the fluid equation reduces to
\begin{equation}\label{eq:app_dn_obi_temp}
  \left[\partial_\tau^2-\frac{2}{3}\frac{\sigma_{pr}^2}{m_b^2 \gamma_b^2} \nabla_\perp^2\right] \delta n_b g(x,y) = 0.
\end{equation}
Considering the slowly varying envelope in $g(x,y)$, the damping of the perturbation amplitude is described by
\begin{equation}
  \left[ \partial_\tau^2 + \frac{2}{3}\frac{\sigma_{pr}^2 k_r^2}{m_b^2\gamma_b^2} \right] \delta n_b = 0.
\end{equation}
The Green's function is readily obtained by applying a Fourier transform for $\tau\geq0$ to
\begin{ngather}
  \delta n_{b,D} = \delta n_b \exp{(-\delta_D \tau)} \\
  \delta_D = \sqrt{\frac{2}{3}}\frac{\sigma_{pr} k_{r}}{\gamma_b m_b}= \sqrt{\frac{2}{3}}\frac{\sigma_{pr}/(m_b c)}{\gamma_b}\frac{k_{r}}{k_p}\omega_p.
\end{ngather}
Consequently, the total growth rate of the bunch perturbation is a sum of the growth rate from the two-stream instability with the damping rate from diffusion, $\Gamma_\mathrm{tot}=\Gamma_\mathrm{TS}\exp{(-\delta_D\tau)}$.

The growth of the two-stream instability is larger for higher wavenumbers, as seen by the spectral parameter $\eta_u$ in \cref{eq:app_nb_green}. However, these wavenumbers are more strongly damped by diffusion. This gives rise to a finite wavenumber $k_{\Gamma_\mathrm{max}}(\tau,\zeta)$ for which the growth is largest, which can be derived in the asymptotic limit by $\partial_{k_r} \exp{(\Gamma_\infty - \delta_D\tau)}/(4\pi\Gamma_\infty)^{1/2}=0$. Further, a cut-off wavenumber $k_\mathrm{cut}(\tau,\zeta)$ exists at which the growth and damping rates are equal, $\exp{(\Gamma_\infty - \delta_D\tau)}/(4\pi\Gamma_\infty)^{1/2}=0$. For higher wavenumbers, an initial bunch perturbation will be damped. Their respective values are numerically evaluated from
\begin{nalign}\label{eq:app_krmax}
  2 \Gamma_\infty(k_{\Gamma_\mathrm{max}}) &= 3(1+k_{\Gamma_\mathrm{max}}^2)\delta_D(k_{\Gamma_\mathrm{max}})\tau +1 \\
  \Gamma_\infty(k_\mathrm{cut}) &= \delta_D(k_\mathrm{cut})\tau+ \ln{\sqrt{4\pi \Gamma_\infty(k_\mathrm{cut})}}
\end{nalign}

\subsection{Transition from TTS to TSI}\label{sec:app_der_tsitts}

\begin{figure}
  \includegraphics[width=0.95\linewidth]{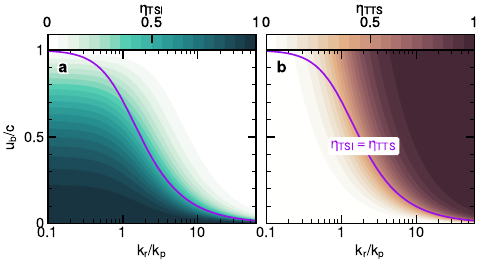}
  \caption{\label{fig:betagr_tts} The spectral dependency $\eta_u$ for a) TSI and b) TTS on bunch velocity $u_b$ and wavenumber $k_r$. The purple line indicates where TSI and TTS equally contribute.}
\end{figure}

To qualitatively compare the dominant regime of the (longitudinal) two-stream and transverse two-stream instability, respectively referred to as \gls{tsi} and \gls{tts}, the spectral parameter from \cref{eq:app_nb_green} can be rewritten to $\eta_u=\eta_\mathrm{TSI}+\eta_\mathrm{TTS}$. The longitudinal and transverse contributions are provided by
\begin{equation}
    \eta_\mathrm{TSI} = \frac{(c^2-u_b^2)k_p^2}{c^2 k_p^2+u_b^2 k_{r}^2}, \quad
    \eta_\mathrm{TTS} = \frac{u_b^2 k_{r}^2}{c^2 k_p^2+u_b^2 k_{r}^2},
\end{equation}
and shown in \cref{fig:betagr_tts}a) and b).

As expected, \gls{tsi} is dominant for non-relativistic bunches, and the longitudinal wakefield component predominantly modulates the bunch. However, for transverse perturbations with a long scale, $k_r/k_p<1$, \gls{tsi} remains dominant even in mildly relativistic regimes. This is a consequence of the transverse electric field scaling $\tilde{\bm{E}}_\perp = \tilde{E_z} \bm{k}_r u_b/(k_p c)$ to the longitudinal field from \cref{eq:app_wake_cos}. \gls{tts} is dominant for highly relativistic bunches or high transverse wavenumbers in mildly relativistic bunches, such that the transverse wakefield predominantly modulates the bunch. Given a negligible energy spread of the bunch, the longitudinal wavenumber of the two-stream instability uniformly equals $k_\zeta=k_e=ck_p/u_b$.
The combined influence of TSI and TTS is generally referred to as oblique instability (OBI) \citep{Bludman1960, Fainberg1969, Thode1976, Califano1998, Bret2010, Chang2016, Shukla2018, Claveria2022}. However, the current filamentation instability (CFI), which becomes dominant for overdense beams, represents a different longitudinal wavenumber ($k_\zeta=0$) and growth scaling as discussed in \citep{Bret2008, Pathak2015}.

\begin{figure}[b]
  \includegraphics[width=0.95\linewidth]{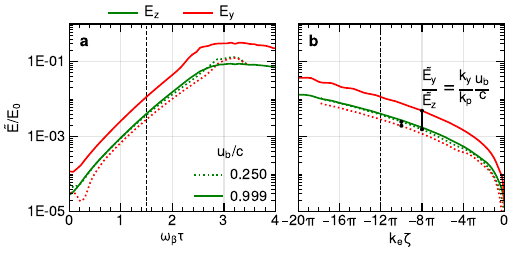}
  \caption{\label{fig:betagr_tts_sim} The longitudinal $E_z$ (green) and transverse $E_y$ (red) electric field envelope for $u_b=0.25\,c$ and $0.999\,c$ as a function of a) propagation time and b) position along the bunch. The solid vertical lines indicate the theoretical ratio between the field components. The dashed lines indicate the respective slice taken.}
\end{figure}

\Cref{fig:betagr_tts_sim} shows the growth of an initial perturbation $k_r/k_p=\pi$ for two different bunch velocities. As can be seen, the growth scales with $\omega_\beta\tau$ and $k_e\tau$, in agreement with \cref{eq:app_ts_growth_asympt}, as the spectral parameter $\eta_u$ remains roughly constant between non-relativistic and relativistic bunches for $k_r/k_p\gtrsim 3$. For a constant wavenumber, $E_y$ is weaker in the non-relativistic limit, given by the theoretical ratio.

Bunches with reduced mass are often used to lower the computational overhead of simulations. It should be noted that the two-stream instability growth scales with $\omega_\beta\sim m_b^{-1/2}$ along the propagation time while damping scales with $\sigma_{pr}/m_b$. Therefore, when scaling the bunch mass, the bunch thermal spread should be scaled by a factor of $(m_b/m_\mathrm{reduced})^{1/2}$ to maintain the ratio of the growth and diffusion rate.

\bibliography{main}{}

\providecommand{\noopsort}[1]{}\providecommand{\singleletter}[1]{#1}%
\begin{thebibliography}{59}%
\makeatletter
\providecommand \@ifxundefined [1]{%
 \@ifx{#1\undefined}
}%
\providecommand \@ifnum [1]{%
 \ifnum #1\expandafter \@firstoftwo
 \else \expandafter \@secondoftwo
 \fi
}%
\providecommand \@ifx [1]{%
 \ifx #1\expandafter \@firstoftwo
 \else \expandafter \@secondoftwo
 \fi
}%
\providecommand \natexlab [1]{#1}%
\providecommand \enquote  [1]{``#1''}%
\providecommand \bibnamefont  [1]{#1}%
\providecommand \bibfnamefont [1]{#1}%
\providecommand \citenamefont [1]{#1}%
\providecommand \href@noop [0]{\@secondoftwo}%
\providecommand \href [0]{\begingroup \@sanitize@url \@href}%
\providecommand \@href[1]{\@@startlink{#1}\@@href}%
\providecommand \@@href[1]{\endgroup#1\@@endlink}%
\providecommand \@sanitize@url [0]{\catcode `\\12\catcode `\$12\catcode
  `\&12\catcode `\#12\catcode `\^12\catcode `\_12\catcode `\%12\relax}%
\providecommand \@@startlink[1]{}%
\providecommand \@@endlink[0]{}%
\providecommand \url  [0]{\begingroup\@sanitize@url \@url }%
\providecommand \@url [1]{\endgroup\@href {#1}{\urlprefix }}%
\providecommand \urlprefix  [0]{URL }%
\providecommand \Eprint [0]{\href }%
\providecommand \doibase [0]{https://doi.org/}%
\providecommand \selectlanguage [0]{\@gobble}%
\providecommand \bibinfo  [0]{\@secondoftwo}%
\providecommand \bibfield  [0]{\@secondoftwo}%
\providecommand \translation [1]{[#1]}%
\providecommand \BibitemOpen [0]{}%
\providecommand \bibitemStop [0]{}%
\providecommand \bibitemNoStop [0]{.\EOS\space}%
\providecommand \EOS [0]{\spacefactor3000\relax}%
\providecommand \BibitemShut  [1]{\csname bibitem#1\endcsname}%
\let\auto@bib@innerbib\@empty
\bibitem [{\citenamefont {Chen}(2016)}]{Chen2016}%
  \BibitemOpen
  \bibfield  {author} {\bibinfo {author} {\bibfnamefont {F.~F.}\ \bibnamefont
  {Chen}},\ }in\ \href {https://doi.org/10.1007/978-3-319-22309-4} {\emph
  {\bibinfo {booktitle} {Introduction to Plasma Physics and Controlled
  Fusion}}}\ (\bibinfo  {publisher} {Springer Interational Publishing
  Switzerland 2016},\ \bibinfo {year} {2016})\ Chap.\ \bibinfo {chapter} {7.3},
  pp.\ \bibinfo {pages} {222--224},\ \bibinfo {edition} {3rd}\ ed.\BibitemShut
  {Stop}%
\bibitem [{\citenamefont {Bret}\ \emph
  {et~al.}(2010{\natexlab{a}})\citenamefont {Bret}, \citenamefont {Gremillet},\
  and\ \citenamefont {Dieckmann}}]{Bret2010}%
  \BibitemOpen
  \bibfield  {author} {\bibinfo {author} {\bibfnamefont {A.}~\bibnamefont
  {Bret}}, \bibinfo {author} {\bibfnamefont {L.}~\bibnamefont {Gremillet}},\
  and\ \bibinfo {author} {\bibfnamefont {M.~E.}\ \bibnamefont {Dieckmann}},\
  }\href {https://doi.org/10.1063/1.3514586} {\bibfield  {journal} {\bibinfo
  {journal} {Physics of Plasmas}\ }\textbf {\bibinfo {volume} {17}},\ \bibinfo
  {pages} {120501} (\bibinfo {year} {2010}{\natexlab{a}})},\ \Eprint
  {https://arxiv.org/abs/https://pubs.aip.org/aip/pop/article-pdf/doi/10.1063/1.3514586/16019035/120501\_1\_online.pdf}
  {https://pubs.aip.org/aip/pop/article-pdf/doi/10.1063/1.3514586/16019035/120501\_1\_online.pdf}
  \BibitemShut {NoStop}%
\bibitem [{\citenamefont {Michno}\ and\ \citenamefont
  {Schlickeiser}(2010)}]{Michno2010}%
  \BibitemOpen
  \bibfield  {author} {\bibinfo {author} {\bibfnamefont {M.~J.}\ \bibnamefont
  {Michno}}\ and\ \bibinfo {author} {\bibfnamefont {R.}~\bibnamefont
  {Schlickeiser}},\ }\href {https://doi.org/10.1088/0004-637X/714/1/868}
  {\bibfield  {journal} {\bibinfo  {journal} {The Astrophysical Journal}\
  }\textbf {\bibinfo {volume} {714}},\ \bibinfo {pages} {868} (\bibinfo {year}
  {2010})}\BibitemShut {NoStop}%
\bibitem [{\citenamefont {Spitkovsky}(2008)}]{Spitkovsky2008}%
  \BibitemOpen
  \bibfield  {author} {\bibinfo {author} {\bibfnamefont {A.}~\bibnamefont
  {Spitkovsky}},\ }\href {https://doi.org/10.1086/590248} {\bibfield  {journal}
  {\bibinfo  {journal} {The Astrophysical Journal}\ }\textbf {\bibinfo {volume}
  {682}},\ \bibinfo {pages} {L5} (\bibinfo {year} {2008})}\BibitemShut
  {NoStop}%
\bibitem [{\citenamefont {Landau}(1946)}]{Landau1946}%
  \BibitemOpen
  \bibfield  {author} {\bibinfo {author} {\bibfnamefont {L.}~\bibnamefont
  {Landau}},\ }\href@noop {} {\bibfield  {journal} {\bibinfo  {journal} {J.
  Phys.}\ }\textbf {\bibinfo {volume} {10}},\ \bibinfo {pages} {25} (\bibinfo
  {year} {1946})}\BibitemShut {NoStop}%
\bibitem [{\citenamefont {Iwamoto}\ \emph {et~al.}(2019)\citenamefont
  {Iwamoto}, \citenamefont {Amano}, \citenamefont {Hoshino}, \citenamefont
  {Matsumoto}, \citenamefont {Niemiec}, \citenamefont {Ligorini}, \citenamefont
  {Kobzar},\ and\ \citenamefont {Pohl}}]{Iwamoto2019}%
  \BibitemOpen
  \bibfield  {author} {\bibinfo {author} {\bibfnamefont {M.}~\bibnamefont
  {Iwamoto}}, \bibinfo {author} {\bibfnamefont {T.}~\bibnamefont {Amano}},
  \bibinfo {author} {\bibfnamefont {M.}~\bibnamefont {Hoshino}}, \bibinfo
  {author} {\bibfnamefont {Y.}~\bibnamefont {Matsumoto}}, \bibinfo {author}
  {\bibfnamefont {J.}~\bibnamefont {Niemiec}}, \bibinfo {author} {\bibfnamefont
  {A.}~\bibnamefont {Ligorini}}, \bibinfo {author} {\bibfnamefont
  {O.}~\bibnamefont {Kobzar}},\ and\ \bibinfo {author} {\bibfnamefont
  {M.}~\bibnamefont {Pohl}},\ }\href {https://doi.org/10.3847/2041-8213/ab4265}
  {\bibfield  {journal} {\bibinfo  {journal} {The Astrophysical Journal
  Letters}\ }\textbf {\bibinfo {volume} {883}},\ \bibinfo {pages} {L35}
  (\bibinfo {year} {2019})}\BibitemShut {NoStop}%
\bibitem [{\citenamefont {Tajima}\ \emph {et~al.}(2020)\citenamefont {Tajima},
  \citenamefont {Yan},\ and\ \citenamefont {Ebisuzaki}}]{Tajima2020}%
  \BibitemOpen
  \bibfield  {author} {\bibinfo {author} {\bibfnamefont {T.}~\bibnamefont
  {Tajima}}, \bibinfo {author} {\bibfnamefont {X.~Q.}\ \bibnamefont {Yan}},\
  and\ \bibinfo {author} {\bibfnamefont {T.}~\bibnamefont {Ebisuzaki}},\
  }\bibfield  {journal} {\bibinfo  {journal} {Reviews of Modern Plasma
  Physics}\ }\textbf {\bibinfo {volume} {4}},\ \href
  {https://doi.org/10.1007/s41614-020-0043-z} {10.1007/s41614-020-0043-z}
  (\bibinfo {year} {2020})\BibitemShut {NoStop}%
\bibitem [{\citenamefont {Hillas}(1984)}]{Hillas1984}%
  \BibitemOpen
  \bibfield  {author} {\bibinfo {author} {\bibfnamefont {A.~M.}\ \bibnamefont
  {Hillas}},\ }\href {https://doi.org/10.1146/annurev.aa.22.090184.002233}
  {\bibfield  {journal} {\bibinfo  {journal} {Annual Review of Astronomy and
  Astrophysics}\ }\textbf {\bibinfo {volume} {22}},\ \bibinfo {pages} {425}
  (\bibinfo {year} {1984})},\ \Eprint
  {https://arxiv.org/abs/https://doi.org/10.1146/annurev.aa.22.090184.002233}
  {https://doi.org/10.1146/annurev.aa.22.090184.002233} \BibitemShut {NoStop}%
\bibitem [{\citenamefont {Piron}(2016)}]{Piron2016}%
  \BibitemOpen
  \bibfield  {author} {\bibinfo {author} {\bibfnamefont {F.}~\bibnamefont
  {Piron}},\ }\href
  {https://doi.org/https://doi.org/10.1016/j.crhy.2016.04.005} {\bibfield
  {journal} {\bibinfo  {journal} {Comptes Rendus Physique}\ }\textbf {\bibinfo
  {volume} {17}},\ \bibinfo {pages} {617} (\bibinfo {year} {2016})},\ \bibinfo
  {note} {gamma-ray astronomy / Astronomie des rayons gamma - Volume
  2}\BibitemShut {NoStop}%
\bibitem [{\citenamefont {Bohdan}\ \emph {et~al.}(2021)\citenamefont {Bohdan},
  \citenamefont {Pohl}, \citenamefont {Niemiec}, \citenamefont {Morris},
  \citenamefont {Matsumoto}, \citenamefont {Amano}, \citenamefont {Hoshino},\
  and\ \citenamefont {Sulaiman}}]{Bohdan2021}%
  \BibitemOpen
  \bibfield  {author} {\bibinfo {author} {\bibfnamefont {A.}~\bibnamefont
  {Bohdan}}, \bibinfo {author} {\bibfnamefont {M.}~\bibnamefont {Pohl}},
  \bibinfo {author} {\bibfnamefont {J.}~\bibnamefont {Niemiec}}, \bibinfo
  {author} {\bibfnamefont {P.~J.}\ \bibnamefont {Morris}}, \bibinfo {author}
  {\bibfnamefont {Y.}~\bibnamefont {Matsumoto}}, \bibinfo {author}
  {\bibfnamefont {T.}~\bibnamefont {Amano}}, \bibinfo {author} {\bibfnamefont
  {M.}~\bibnamefont {Hoshino}},\ and\ \bibinfo {author} {\bibfnamefont
  {A.}~\bibnamefont {Sulaiman}},\ }\href
  {https://doi.org/10.1103/PhysRevLett.126.095101} {\bibfield  {journal}
  {\bibinfo  {journal} {Phys. Rev. Lett.}\ }\textbf {\bibinfo {volume} {126}},\
  \bibinfo {pages} {095101} (\bibinfo {year} {2021})}\BibitemShut {NoStop}%
\bibitem [{\citenamefont {Medvedev}\ and\ \citenamefont
  {Loeb}(1999)}]{Medvedev1999}%
  \BibitemOpen
  \bibfield  {author} {\bibinfo {author} {\bibfnamefont {M.~V.}\ \bibnamefont
  {Medvedev}}\ and\ \bibinfo {author} {\bibfnamefont {A.}~\bibnamefont
  {Loeb}},\ }\href {https://doi.org/10.1086/308038} {\bibfield  {journal}
  {\bibinfo  {journal} {The Astrophysical Journal}\ }\textbf {\bibinfo {volume}
  {526}},\ \bibinfo {pages} {697} (\bibinfo {year} {1999})}\BibitemShut
  {NoStop}%
\bibitem [{\citenamefont {Perna}\ \emph {et~al.}(2016)\citenamefont {Perna},
  \citenamefont {Lazzati},\ and\ \citenamefont {Giacomazzo}}]{Perna2016}%
  \BibitemOpen
  \bibfield  {author} {\bibinfo {author} {\bibfnamefont {R.}~\bibnamefont
  {Perna}}, \bibinfo {author} {\bibfnamefont {D.}~\bibnamefont {Lazzati}},\
  and\ \bibinfo {author} {\bibfnamefont {B.}~\bibnamefont {Giacomazzo}},\
  }\href {https://doi.org/10.3847/2041-8205/821/1/L18} {\bibfield  {journal}
  {\bibinfo  {journal} {The Astrophysical Journal Letters}\ }\textbf {\bibinfo
  {volume} {821}},\ \bibinfo {pages} {L18} (\bibinfo {year}
  {2016})}\BibitemShut {NoStop}%
\bibitem [{\citenamefont {Fiuza}\ \emph {et~al.}(2020)\citenamefont {Fiuza},
  \citenamefont {Swadling}, \citenamefont {Grassi},\ and\ \citenamefont
  {et~al.}}]{Fiuza2020}%
  \BibitemOpen
  \bibfield  {author} {\bibinfo {author} {\bibfnamefont {F.}~\bibnamefont
  {Fiuza}}, \bibinfo {author} {\bibfnamefont {G.}~\bibnamefont {Swadling}},
  \bibinfo {author} {\bibfnamefont {A.}~\bibnamefont {Grassi}},\ and\ \bibinfo
  {author} {\bibnamefont {et~al.}},\ }\href
  {https://doi.org/10.1038/s41567-020-0919-4} {\bibfield  {journal} {\bibinfo
  {journal} {Nat. Phys.}\ }\textbf {\bibinfo {volume} {16}},\ \bibinfo {pages}
  {916} (\bibinfo {year} {2020})}\BibitemShut {NoStop}%
\bibitem [{\citenamefont {Arrowsmith}\ \emph {et~al.}(2021)\citenamefont
  {Arrowsmith}, \citenamefont {Shukla}, \citenamefont {Charitonidis},
  \citenamefont {Boni}, \citenamefont {Chen}, \citenamefont {Davenne},
  \citenamefont {Dyson}, \citenamefont {Froula}, \citenamefont {Gudmundsson},
  \citenamefont {Huffman}, \citenamefont {Kadi}, \citenamefont {Reville},
  \citenamefont {Richardson}, \citenamefont {Sarkar}, \citenamefont {Shaw},
  \citenamefont {Silva}, \citenamefont {Simon}, \citenamefont {Trines},
  \citenamefont {Bingham},\ and\ \citenamefont {Gregori}}]{Arrowsmith2021}%
  \BibitemOpen
  \bibfield  {author} {\bibinfo {author} {\bibfnamefont {C.~D.}\ \bibnamefont
  {Arrowsmith}}, \bibinfo {author} {\bibfnamefont {N.}~\bibnamefont {Shukla}},
  \bibinfo {author} {\bibfnamefont {N.}~\bibnamefont {Charitonidis}}, \bibinfo
  {author} {\bibfnamefont {R.}~\bibnamefont {Boni}}, \bibinfo {author}
  {\bibfnamefont {H.}~\bibnamefont {Chen}}, \bibinfo {author} {\bibfnamefont
  {T.}~\bibnamefont {Davenne}}, \bibinfo {author} {\bibfnamefont
  {A.}~\bibnamefont {Dyson}}, \bibinfo {author} {\bibfnamefont {D.~H.}\
  \bibnamefont {Froula}}, \bibinfo {author} {\bibfnamefont {J.~T.}\
  \bibnamefont {Gudmundsson}}, \bibinfo {author} {\bibfnamefont {B.~T.}\
  \bibnamefont {Huffman}}, \bibinfo {author} {\bibfnamefont {Y.}~\bibnamefont
  {Kadi}}, \bibinfo {author} {\bibfnamefont {B.}~\bibnamefont {Reville}},
  \bibinfo {author} {\bibfnamefont {S.}~\bibnamefont {Richardson}}, \bibinfo
  {author} {\bibfnamefont {S.}~\bibnamefont {Sarkar}}, \bibinfo {author}
  {\bibfnamefont {J.~L.}\ \bibnamefont {Shaw}}, \bibinfo {author}
  {\bibfnamefont {L.~O.}\ \bibnamefont {Silva}}, \bibinfo {author}
  {\bibfnamefont {P.}~\bibnamefont {Simon}}, \bibinfo {author} {\bibfnamefont
  {R.~M. G.~M.}\ \bibnamefont {Trines}}, \bibinfo {author} {\bibfnamefont
  {R.}~\bibnamefont {Bingham}},\ and\ \bibinfo {author} {\bibfnamefont
  {G.}~\bibnamefont {Gregori}},\ }\href
  {https://doi.org/10.1103/PhysRevResearch.3.023103} {\bibfield  {journal}
  {\bibinfo  {journal} {Phys. Rev. Res.}\ }\textbf {\bibinfo {volume} {3}},\
  \bibinfo {pages} {023103} (\bibinfo {year} {2021})}\BibitemShut {NoStop}%
\bibitem [{\citenamefont {Zhang}\ \emph {et~al.}(2022)\citenamefont {Zhang},
  \citenamefont {Wu}, \citenamefont {Sinclair}, \citenamefont {Farrell},
  \citenamefont {Marsh}, \citenamefont {Petrushina}, \citenamefont
  {Vafaei-Najafabadi}, \citenamefont {Gaikwad}, \citenamefont {Kupfer},
  \citenamefont {Kusche}, \citenamefont {Fedurin}, \citenamefont {Pogorelsky},
  \citenamefont {Polyanskiy}, \citenamefont {Huang}, \citenamefont {Hua},
  \citenamefont {Lu}, \citenamefont {Mori},\ and\ \citenamefont
  {Joshi}}]{Zhang2022}%
  \BibitemOpen
  \bibfield  {author} {\bibinfo {author} {\bibfnamefont {C.}~\bibnamefont
  {Zhang}}, \bibinfo {author} {\bibfnamefont {Y.}~\bibnamefont {Wu}}, \bibinfo
  {author} {\bibfnamefont {M.}~\bibnamefont {Sinclair}}, \bibinfo {author}
  {\bibfnamefont {A.}~\bibnamefont {Farrell}}, \bibinfo {author} {\bibfnamefont
  {K.~A.}\ \bibnamefont {Marsh}}, \bibinfo {author} {\bibfnamefont
  {I.}~\bibnamefont {Petrushina}}, \bibinfo {author} {\bibfnamefont
  {N.}~\bibnamefont {Vafaei-Najafabadi}}, \bibinfo {author} {\bibfnamefont
  {A.}~\bibnamefont {Gaikwad}}, \bibinfo {author} {\bibfnamefont
  {R.}~\bibnamefont {Kupfer}}, \bibinfo {author} {\bibfnamefont
  {K.}~\bibnamefont {Kusche}}, \bibinfo {author} {\bibfnamefont
  {M.}~\bibnamefont {Fedurin}}, \bibinfo {author} {\bibfnamefont
  {I.}~\bibnamefont {Pogorelsky}}, \bibinfo {author} {\bibfnamefont
  {M.}~\bibnamefont {Polyanskiy}}, \bibinfo {author} {\bibfnamefont {C.-K.}\
  \bibnamefont {Huang}}, \bibinfo {author} {\bibfnamefont {J.}~\bibnamefont
  {Hua}}, \bibinfo {author} {\bibfnamefont {W.}~\bibnamefont {Lu}}, \bibinfo
  {author} {\bibfnamefont {W.~B.}\ \bibnamefont {Mori}},\ and\ \bibinfo
  {author} {\bibfnamefont {C.}~\bibnamefont {Joshi}},\ }\href
  {https://doi.org/10.1073/pnas.2211713119} {\bibfield  {journal} {\bibinfo
  {journal} {Proceedings of the National Academy of Sciences}\ }\textbf
  {\bibinfo {volume} {119}},\ \bibinfo {pages} {e2211713119} (\bibinfo {year}
  {2022})},\ \Eprint
  {https://arxiv.org/abs/https://www.pnas.org/doi/pdf/10.1073/pnas.2211713119}
  {https://www.pnas.org/doi/pdf/10.1073/pnas.2211713119} \BibitemShut {NoStop}%
\bibitem [{\citenamefont {Chen}\ \emph {et~al.}(1985)\citenamefont {Chen},
  \citenamefont {Dawson}, \citenamefont {Huff},\ and\ \citenamefont
  {Katsouleas}}]{Chen1985}%
  \BibitemOpen
  \bibfield  {author} {\bibinfo {author} {\bibfnamefont {P.}~\bibnamefont
  {Chen}}, \bibinfo {author} {\bibfnamefont {J.~M.}\ \bibnamefont {Dawson}},
  \bibinfo {author} {\bibfnamefont {R.~W.}\ \bibnamefont {Huff}},\ and\
  \bibinfo {author} {\bibfnamefont {T.}~\bibnamefont {Katsouleas}},\ }\href
  {https://doi.org/10.1103/PhysRevLett.54.693} {\bibfield  {journal} {\bibinfo
  {journal} {Phys. Rev. Lett.}\ }\textbf {\bibinfo {volume} {54}},\ \bibinfo
  {pages} {693} (\bibinfo {year} {1985})}\BibitemShut {NoStop}%
\bibitem [{\citenamefont {Macchi}\ and\ \citenamefont
  {Pegoraro}(2018)}]{Macchi2018}%
  \BibitemOpen
  \bibfield  {author} {\bibinfo {author} {\bibfnamefont {A.}~\bibnamefont
  {Macchi}}\ and\ \bibinfo {author} {\bibfnamefont {F.}~\bibnamefont
  {Pegoraro}},\ }\href {https://doi.org/10.1038/s41566-018-0181-9} {\bibfield
  {journal} {\bibinfo  {journal} {Nature Photonics}\ }\textbf {\bibinfo
  {volume} {12}},\ \bibinfo {pages} {314} (\bibinfo {year} {2018})}\BibitemShut
  {NoStop}%
\bibitem [{\citenamefont {Schroeder}\ \emph {et~al.}(2011)\citenamefont
  {Schroeder}, \citenamefont {Benedetti}, \citenamefont {Esarey}, \citenamefont
  {Gr\"uner},\ and\ \citenamefont {Leemans}}]{Schroeder2011}%
  \BibitemOpen
  \bibfield  {author} {\bibinfo {author} {\bibfnamefont {C.~B.}\ \bibnamefont
  {Schroeder}}, \bibinfo {author} {\bibfnamefont {C.}~\bibnamefont
  {Benedetti}}, \bibinfo {author} {\bibfnamefont {E.}~\bibnamefont {Esarey}},
  \bibinfo {author} {\bibfnamefont {F.~J.}\ \bibnamefont {Gr\"uner}},\ and\
  \bibinfo {author} {\bibfnamefont {W.~P.}\ \bibnamefont {Leemans}},\ }\href
  {https://doi.org/10.1103/PhysRevLett.107.145002} {\bibfield  {journal}
  {\bibinfo  {journal} {Phys. Rev. Lett.}\ }\textbf {\bibinfo {volume} {107}},\
  \bibinfo {pages} {145002} (\bibinfo {year} {2011})}\BibitemShut {NoStop}%
\bibitem [{\citenamefont {Caldwell}\ and\ \citenamefont
  {Lotov}(2011)}]{Caldwell2011}%
  \BibitemOpen
  \bibfield  {author} {\bibinfo {author} {\bibfnamefont {A.}~\bibnamefont
  {Caldwell}}\ and\ \bibinfo {author} {\bibfnamefont {K.~V.}\ \bibnamefont
  {Lotov}},\ }\href {https://doi.org/10.1063/1.3641973} {\bibfield  {journal}
  {\bibinfo  {journal} {Physics of Plasmas}\ }\textbf {\bibinfo {volume}
  {18}},\ \bibinfo {pages} {103101} (\bibinfo {year} {2011})},\ \Eprint
  {https://arxiv.org/abs/https://pubs.aip.org/aip/pop/article-pdf/doi/10.1063/1.3641973/13613329/103101\_1\_online.pdf}
  {https://pubs.aip.org/aip/pop/article-pdf/doi/10.1063/1.3641973/13613329/103101\_1\_online.pdf}
  \BibitemShut {NoStop}%
\bibitem [{\citenamefont {{AWAKE Collaboration}}(2018)}]{Adli2018}%
  \BibitemOpen
  \bibfield  {author} {\bibinfo {author} {\bibnamefont {{AWAKE
  Collaboration}}},\ }\href {https://doi.org/10.1038/s41586-018-0485-4}
  {\bibfield  {journal} {\bibinfo  {journal} {Nature}\ }\textbf {\bibinfo
  {volume} {561}},\ \bibinfo {pages} {363} (\bibinfo {year}
  {2018})}\BibitemShut {NoStop}%
\bibitem [{\citenamefont {Verra}\ \emph {et~al.}(2024)\citenamefont {Verra},
  \citenamefont {Muggli} \emph {et~al.}}]{Verra2024}%
  \BibitemOpen
  \bibfield  {author} {\bibinfo {author} {\bibfnamefont {L.}~\bibnamefont
  {Verra}}, \bibinfo {author} {\bibfnamefont {P.}~\bibnamefont {Muggli}}, \emph
  {et~al.} (\bibinfo {collaboration} {AWAKE Collaboration}),\ }\href
  {https://doi.org/10.1103/PhysRevE.109.055203} {\bibfield  {journal} {\bibinfo
   {journal} {Phys. Rev. E}\ }\textbf {\bibinfo {volume} {109}},\ \bibinfo
  {pages} {055203} (\bibinfo {year} {2024})}\BibitemShut {NoStop}%
\bibitem [{\citenamefont {Allen}\ \emph {et~al.}(2012)\citenamefont {Allen},
  \citenamefont {Yakimenko}, \citenamefont {Babzien}, \citenamefont {Fedurin},
  \citenamefont {Kusche},\ and\ \citenamefont {Muggli}}]{Allen2012}%
  \BibitemOpen
  \bibfield  {author} {\bibinfo {author} {\bibfnamefont {B.}~\bibnamefont
  {Allen}}, \bibinfo {author} {\bibfnamefont {V.}~\bibnamefont {Yakimenko}},
  \bibinfo {author} {\bibfnamefont {M.}~\bibnamefont {Babzien}}, \bibinfo
  {author} {\bibfnamefont {M.}~\bibnamefont {Fedurin}}, \bibinfo {author}
  {\bibfnamefont {K.}~\bibnamefont {Kusche}},\ and\ \bibinfo {author}
  {\bibfnamefont {P.}~\bibnamefont {Muggli}},\ }\href
  {https://doi.org/10.1103/PhysRevLett.109.185007} {\bibfield  {journal}
  {\bibinfo  {journal} {Phys. Rev. Lett.}\ }\textbf {\bibinfo {volume} {109}},\
  \bibinfo {pages} {185007} (\bibinfo {year} {2012})}\BibitemShut {NoStop}%
\bibitem [{\citenamefont {Weibel}(1959)}]{Weibel1959}%
  \BibitemOpen
  \bibfield  {author} {\bibinfo {author} {\bibfnamefont {E.~S.}\ \bibnamefont
  {Weibel}},\ }\href {https://doi.org/10.1103/PhysRevLett.2.83} {\bibfield
  {journal} {\bibinfo  {journal} {Phys. Rev. Lett.}\ }\textbf {\bibinfo
  {volume} {2}},\ \bibinfo {pages} {83} (\bibinfo {year} {1959})}\BibitemShut
  {NoStop}%
\bibitem [{\citenamefont {Fried}(1959)}]{Fried1959}%
  \BibitemOpen
  \bibfield  {author} {\bibinfo {author} {\bibfnamefont {B.~D.}\ \bibnamefont
  {Fried}},\ }\href {https://doi.org/10.1063/1.1705933} {\bibfield  {journal}
  {\bibinfo  {journal} {The Physics of Fluids}\ }\textbf {\bibinfo {volume}
  {2}},\ \bibinfo {pages} {337} (\bibinfo {year} {1959})},\ \Eprint
  {https://arxiv.org/abs/https://pubs.aip.org/aip/pfl/article-pdf/2/3/337/12401439/337\_1\_online.pdf}
  {https://pubs.aip.org/aip/pfl/article-pdf/2/3/337/12401439/337\_1\_online.pdf}
  \BibitemShut {NoStop}%
\bibitem [{\citenamefont {Bludman}\ \emph {et~al.}(1960)\citenamefont
  {Bludman}, \citenamefont {Watson},\ and\ \citenamefont
  {Rosenbluth}}]{Bludman1960}%
  \BibitemOpen
  \bibfield  {author} {\bibinfo {author} {\bibfnamefont {S.~A.}\ \bibnamefont
  {Bludman}}, \bibinfo {author} {\bibfnamefont {K.~M.}\ \bibnamefont
  {Watson}},\ and\ \bibinfo {author} {\bibfnamefont {M.~N.}\ \bibnamefont
  {Rosenbluth}},\ }\href {https://doi.org/10.1063/1.1706121} {\bibfield
  {journal} {\bibinfo  {journal} {The Physics of Fluids}\ }\textbf {\bibinfo
  {volume} {3}},\ \bibinfo {pages} {747} (\bibinfo {year} {1960})},\ \Eprint
  {https://arxiv.org/abs/https://pubs.aip.org/aip/pfl/article-pdf/3/5/747/12499127/747\_1\_online.pdf}
  {https://pubs.aip.org/aip/pfl/article-pdf/3/5/747/12499127/747\_1\_online.pdf}
  \BibitemShut {NoStop}%
\bibitem [{\citenamefont {{Fa{\v{i}}nberg}}\ \emph {et~al.}(1969)\citenamefont
  {{Fa{\v{i}}nberg}}, \citenamefont {{Shapiro}},\ and\ \citenamefont
  {{Shevchenko}}}]{Fainberg1969}%
  \BibitemOpen
  \bibfield  {author} {\bibinfo {author} {\bibfnamefont {Y.~B.}\ \bibnamefont
  {{Fa{\v{i}}nberg}}}, \bibinfo {author} {\bibfnamefont {V.~D.}\ \bibnamefont
  {{Shapiro}}},\ and\ \bibinfo {author} {\bibfnamefont {V.~I.}\ \bibnamefont
  {{Shevchenko}}},\ }\href@noop {} {\bibfield  {journal} {\bibinfo  {journal}
  {Soviet Journal of Experimental and Theoretical Physics}\ }\textbf {\bibinfo
  {volume} {30}},\ \bibinfo {pages} {528} (\bibinfo {year} {1969})}\BibitemShut
  {NoStop}%
\bibitem [{\citenamefont {Bret}\ \emph {et~al.}(2004)\citenamefont {Bret},
  \citenamefont {Firpo},\ and\ \citenamefont {Deutsch}}]{Bret2004}%
  \BibitemOpen
  \bibfield  {author} {\bibinfo {author} {\bibfnamefont {A.}~\bibnamefont
  {Bret}}, \bibinfo {author} {\bibfnamefont {M.-C.}\ \bibnamefont {Firpo}},\
  and\ \bibinfo {author} {\bibfnamefont {C.}~\bibnamefont {Deutsch}},\ }\href
  {https://doi.org/10.1103/PhysRevE.70.046401} {\bibfield  {journal} {\bibinfo
  {journal} {Phys. Rev. E}\ }\textbf {\bibinfo {volume} {70}},\ \bibinfo
  {pages} {046401} (\bibinfo {year} {2004})}\BibitemShut {NoStop}%
\bibitem [{\citenamefont {Tonks}\ and\ \citenamefont
  {Langmuir}(1929)}]{Langmuir1929}%
  \BibitemOpen
  \bibfield  {author} {\bibinfo {author} {\bibfnamefont {L.}~\bibnamefont
  {Tonks}}\ and\ \bibinfo {author} {\bibfnamefont {I.}~\bibnamefont
  {Langmuir}},\ }\href {https://doi.org/10.1103/PhysRev.33.195} {\bibfield
  {journal} {\bibinfo  {journal} {Phys. Rev.}\ }\textbf {\bibinfo {volume}
  {33}},\ \bibinfo {pages} {195} (\bibinfo {year} {1929})}\BibitemShut
  {NoStop}%
\bibitem [{\citenamefont {Dawson}(1959)}]{Dawson1959}%
  \BibitemOpen
  \bibfield  {author} {\bibinfo {author} {\bibfnamefont {J.~M.}\ \bibnamefont
  {Dawson}},\ }\href {https://doi.org/10.1103/PhysRev.113.383} {\bibfield
  {journal} {\bibinfo  {journal} {Phys. Rev.}\ }\textbf {\bibinfo {volume}
  {113}},\ \bibinfo {pages} {383} (\bibinfo {year} {1959})}\BibitemShut
  {NoStop}%
\bibitem [{\citenamefont {Lawson}\ and\ \citenamefont
  {Lawson}(1977)}]{Lawson1977}%
  \BibitemOpen
  \bibfield  {author} {\bibinfo {author} {\bibfnamefont {J.}~\bibnamefont
  {Lawson}}\ and\ \bibinfo {author} {\bibfnamefont {J.}~\bibnamefont
  {Lawson}},\ }\href {https://books.google.de/books?id=g9l8AAAAIAAJ} {\emph
  {\bibinfo {title} {The Physics of Charged-particle Beams}}},\ International
  series of monographs on physics\ (\bibinfo  {publisher} {Clarendon Press},\
  \bibinfo {year} {1977})\BibitemShut {NoStop}%
\bibitem [{\citenamefont {Watson}\ \emph {et~al.}(1960)\citenamefont {Watson},
  \citenamefont {Bludman},\ and\ \citenamefont {Rosenbluth}}]{Watson1960}%
  \BibitemOpen
  \bibfield  {author} {\bibinfo {author} {\bibfnamefont {K.~M.}\ \bibnamefont
  {Watson}}, \bibinfo {author} {\bibfnamefont {S.~A.}\ \bibnamefont
  {Bludman}},\ and\ \bibinfo {author} {\bibfnamefont {M.~N.}\ \bibnamefont
  {Rosenbluth}},\ }\href {https://doi.org/10.1063/1.1706120} {\bibfield
  {journal} {\bibinfo  {journal} {The Physics of Fluids}\ }\textbf {\bibinfo
  {volume} {3}},\ \bibinfo {pages} {741} (\bibinfo {year} {1960})},\ \Eprint
  {https://arxiv.org/abs/https://pubs.aip.org/aip/pfl/article-pdf/3/5/741/12499098/741\_1\_online.pdf}
  {https://pubs.aip.org/aip/pfl/article-pdf/3/5/741/12499098/741\_1\_online.pdf}
  \BibitemShut {NoStop}%
\bibitem [{\citenamefont {Davidson}\ \emph {et~al.}(1972)\citenamefont
  {Davidson}, \citenamefont {Hammer}, \citenamefont {Haber},\ and\
  \citenamefont {Wagner}}]{Davidson1972}%
  \BibitemOpen
  \bibfield  {author} {\bibinfo {author} {\bibfnamefont {R.~C.}\ \bibnamefont
  {Davidson}}, \bibinfo {author} {\bibfnamefont {D.~A.}\ \bibnamefont
  {Hammer}}, \bibinfo {author} {\bibfnamefont {I.}~\bibnamefont {Haber}},\ and\
  \bibinfo {author} {\bibfnamefont {C.~E.}\ \bibnamefont {Wagner}},\ }\href
  {https://doi.org/10.1063/1.1693910} {\bibfield  {journal} {\bibinfo
  {journal} {The Physics of Fluids}\ }\textbf {\bibinfo {volume} {15}},\
  \bibinfo {pages} {317} (\bibinfo {year} {1972})},\ \Eprint
  {https://arxiv.org/abs/https://pubs.aip.org/aip/pfl/article-pdf/15/2/317/12743024/317\_1\_online.pdf}
  {https://pubs.aip.org/aip/pfl/article-pdf/15/2/317/12743024/317\_1\_online.pdf}
  \BibitemShut {NoStop}%
\bibitem [{\citenamefont {Silva}\ \emph {et~al.}(2002)\citenamefont {Silva},
  \citenamefont {Fonseca}, \citenamefont {Tonge}, \citenamefont {Mori},\ and\
  \citenamefont {Dawson}}]{Silva2002}%
  \BibitemOpen
  \bibfield  {author} {\bibinfo {author} {\bibfnamefont {L.~O.}\ \bibnamefont
  {Silva}}, \bibinfo {author} {\bibfnamefont {R.~A.}\ \bibnamefont {Fonseca}},
  \bibinfo {author} {\bibfnamefont {J.~W.}\ \bibnamefont {Tonge}}, \bibinfo
  {author} {\bibfnamefont {W.~B.}\ \bibnamefont {Mori}},\ and\ \bibinfo
  {author} {\bibfnamefont {J.~M.}\ \bibnamefont {Dawson}},\ }\href
  {https://doi.org/10.1063/1.1476004} {\bibfield  {journal} {\bibinfo
  {journal} {Physics of Plasmas}\ }\textbf {\bibinfo {volume} {9}},\ \bibinfo
  {pages} {2458} (\bibinfo {year} {2002})},\ \Eprint
  {https://arxiv.org/abs/https://pubs.aip.org/aip/pop/article-pdf/9/6/2458/19097722/2458\_1\_online.pdf}
  {https://pubs.aip.org/aip/pop/article-pdf/9/6/2458/19097722/2458\_1\_online.pdf}
  \BibitemShut {NoStop}%
\bibitem [{\citenamefont {Jia}\ \emph {et~al.}(2013)\citenamefont {Jia},
  \citenamefont {Cai}, \citenamefont {Wang}, \citenamefont {Zhu}, \citenamefont
  {Sheng},\ and\ \citenamefont {He}}]{Jia2013}%
  \BibitemOpen
  \bibfield  {author} {\bibinfo {author} {\bibfnamefont {Q.}~\bibnamefont
  {Jia}}, \bibinfo {author} {\bibfnamefont {H.-b.}\ \bibnamefont {Cai}},
  \bibinfo {author} {\bibfnamefont {W.-w.}\ \bibnamefont {Wang}}, \bibinfo
  {author} {\bibfnamefont {S.-p.}\ \bibnamefont {Zhu}}, \bibinfo {author}
  {\bibfnamefont {Z.~M.}\ \bibnamefont {Sheng}},\ and\ \bibinfo {author}
  {\bibfnamefont {X.~T.}\ \bibnamefont {He}},\ }\href
  {https://doi.org/10.1063/1.4796052} {\bibfield  {journal} {\bibinfo
  {journal} {Physics of Plasmas}\ }\textbf {\bibinfo {volume} {20}},\ \bibinfo
  {pages} {032113} (\bibinfo {year} {2013})},\ \Eprint
  {https://arxiv.org/abs/https://pubs.aip.org/aip/pop/article-pdf/doi/10.1063/1.4796052/14795548/032113\_1\_online.pdf}
  {https://pubs.aip.org/aip/pop/article-pdf/doi/10.1063/1.4796052/14795548/032113\_1\_online.pdf}
  \BibitemShut {NoStop}%
\bibitem [{\citenamefont {Pathak}\ \emph {et~al.}(2015)\citenamefont {Pathak},
  \citenamefont {Grismayer}, \citenamefont {Stockem}, \citenamefont {Fonseca},\
  and\ \citenamefont {Silva}}]{Pathak2015}%
  \BibitemOpen
  \bibfield  {author} {\bibinfo {author} {\bibfnamefont {V.~B.}\ \bibnamefont
  {Pathak}}, \bibinfo {author} {\bibfnamefont {T.}~\bibnamefont {Grismayer}},
  \bibinfo {author} {\bibfnamefont {A.}~\bibnamefont {Stockem}}, \bibinfo
  {author} {\bibfnamefont {R.~A.}\ \bibnamefont {Fonseca}},\ and\ \bibinfo
  {author} {\bibfnamefont {L.~O.}\ \bibnamefont {Silva}},\ }\href
  {https://doi.org/10.1088/1367-2630/17/4/043049} {\bibfield  {journal}
  {\bibinfo  {journal} {New Journal of Physics}\ }\textbf {\bibinfo {volume}
  {17}},\ \bibinfo {pages} {043049} (\bibinfo {year} {2015})}\BibitemShut
  {NoStop}%
\bibitem [{\citenamefont {Bret}\ \emph
  {et~al.}(2010{\natexlab{b}})\citenamefont {Bret}, \citenamefont {Gremillet},\
  and\ \citenamefont {B\'enisti}}]{Bret2010b}%
  \BibitemOpen
  \bibfield  {author} {\bibinfo {author} {\bibfnamefont {A.}~\bibnamefont
  {Bret}}, \bibinfo {author} {\bibfnamefont {L.}~\bibnamefont {Gremillet}},\
  and\ \bibinfo {author} {\bibfnamefont {D.}~\bibnamefont {B\'enisti}},\ }\href
  {https://doi.org/10.1103/PhysRevE.81.036402} {\bibfield  {journal} {\bibinfo
  {journal} {Phys. Rev. E}\ }\textbf {\bibinfo {volume} {81}},\ \bibinfo
  {pages} {036402} (\bibinfo {year} {2010}{\natexlab{b}})}\BibitemShut
  {NoStop}%
\bibitem [{\citenamefont {Bers}(1983)}]{Bers1983}%
  \BibitemOpen
  \bibfield  {author} {\bibinfo {author} {\bibfnamefont {A.}~\bibnamefont
  {Bers}},\ }\bibinfo {title} {Handbook of plasma physics. vol. 1}\ (\bibinfo
  {publisher} {North-Holland Publ.},\ \bibinfo {address} {Amsterdam},\ \bibinfo
  {year} {1983})\ Chap.\ \bibinfo {chapter} {3.2 Space-time evolution of plasma
  instabilities - Absolute and convective}, pp.\ \bibinfo {pages}
  {451--517}\BibitemShut {NoStop}%
\bibitem [{\citenamefont {Jones}\ \emph {et~al.}(1983)\citenamefont {Jones},
  \citenamefont {Lemons},\ and\ \citenamefont {Mostrom}}]{Jones1983}%
  \BibitemOpen
  \bibfield  {author} {\bibinfo {author} {\bibfnamefont {M.~E.}\ \bibnamefont
  {Jones}}, \bibinfo {author} {\bibfnamefont {D.~S.}\ \bibnamefont {Lemons}},\
  and\ \bibinfo {author} {\bibfnamefont {M.~A.}\ \bibnamefont {Mostrom}},\
  }\href {https://doi.org/10.1063/1.864044} {\bibfield  {journal} {\bibinfo
  {journal} {The Physics of Fluids}\ }\textbf {\bibinfo {volume} {26}},\
  \bibinfo {pages} {2784} (\bibinfo {year} {1983})},\ \Eprint
  {https://arxiv.org/abs/https://pubs.aip.org/aip/pfl/article-pdf/26/10/2784/12495716/2784\_1\_online.pdf}
  {https://pubs.aip.org/aip/pfl/article-pdf/26/10/2784/12495716/2784\_1\_online.pdf}
  \BibitemShut {NoStop}%
\bibitem [{\citenamefont {San Miguel~Claveria}\ \emph
  {et~al.}(2022)\citenamefont {San Miguel~Claveria}, \citenamefont {Davoine},
  \citenamefont {Peterson}, \citenamefont {Gilljohann}, \citenamefont
  {Andriyash}, \citenamefont {Ariniello}, \citenamefont {Clarke}, \citenamefont
  {Ekerfelt}, \citenamefont {Emma}, \citenamefont {Faure}, \citenamefont
  {Gessner}, \citenamefont {Hogan}, \citenamefont {Joshi}, \citenamefont
  {Keitel}, \citenamefont {Knetsch}, \citenamefont {Kononenko}, \citenamefont
  {Litos}, \citenamefont {Mankovska}, \citenamefont {Marsh}, \citenamefont
  {Matheron}, \citenamefont {Nie}, \citenamefont {O'Shea}, \citenamefont
  {Storey}, \citenamefont {Vafaei-Najafabadi}, \citenamefont {Wu},
  \citenamefont {Xu}, \citenamefont {Yan}, \citenamefont {Zhang}, \citenamefont
  {Tamburini}, \citenamefont {Fiuza}, \citenamefont {Gremillet},\ and\
  \citenamefont {Corde}}]{Claveria2022}%
  \BibitemOpen
  \bibfield  {author} {\bibinfo {author} {\bibfnamefont {P.}~\bibnamefont {San
  Miguel~Claveria}}, \bibinfo {author} {\bibfnamefont {X.}~\bibnamefont
  {Davoine}}, \bibinfo {author} {\bibfnamefont {J.~R.}\ \bibnamefont
  {Peterson}}, \bibinfo {author} {\bibfnamefont {M.}~\bibnamefont
  {Gilljohann}}, \bibinfo {author} {\bibfnamefont {I.}~\bibnamefont
  {Andriyash}}, \bibinfo {author} {\bibfnamefont {R.}~\bibnamefont
  {Ariniello}}, \bibinfo {author} {\bibfnamefont {C.}~\bibnamefont {Clarke}},
  \bibinfo {author} {\bibfnamefont {H.}~\bibnamefont {Ekerfelt}}, \bibinfo
  {author} {\bibfnamefont {C.}~\bibnamefont {Emma}}, \bibinfo {author}
  {\bibfnamefont {J.}~\bibnamefont {Faure}}, \bibinfo {author} {\bibfnamefont
  {S.}~\bibnamefont {Gessner}}, \bibinfo {author} {\bibfnamefont {M.~J.}\
  \bibnamefont {Hogan}}, \bibinfo {author} {\bibfnamefont {C.}~\bibnamefont
  {Joshi}}, \bibinfo {author} {\bibfnamefont {C.~H.}\ \bibnamefont {Keitel}},
  \bibinfo {author} {\bibfnamefont {A.}~\bibnamefont {Knetsch}}, \bibinfo
  {author} {\bibfnamefont {O.}~\bibnamefont {Kononenko}}, \bibinfo {author}
  {\bibfnamefont {M.}~\bibnamefont {Litos}}, \bibinfo {author} {\bibfnamefont
  {Y.}~\bibnamefont {Mankovska}}, \bibinfo {author} {\bibfnamefont
  {K.}~\bibnamefont {Marsh}}, \bibinfo {author} {\bibfnamefont
  {A.}~\bibnamefont {Matheron}}, \bibinfo {author} {\bibfnamefont
  {Z.}~\bibnamefont {Nie}}, \bibinfo {author} {\bibfnamefont {B.}~\bibnamefont
  {O'Shea}}, \bibinfo {author} {\bibfnamefont {D.}~\bibnamefont {Storey}},
  \bibinfo {author} {\bibfnamefont {N.}~\bibnamefont {Vafaei-Najafabadi}},
  \bibinfo {author} {\bibfnamefont {Y.}~\bibnamefont {Wu}}, \bibinfo {author}
  {\bibfnamefont {X.}~\bibnamefont {Xu}}, \bibinfo {author} {\bibfnamefont
  {J.}~\bibnamefont {Yan}}, \bibinfo {author} {\bibfnamefont {C.}~\bibnamefont
  {Zhang}}, \bibinfo {author} {\bibfnamefont {M.}~\bibnamefont {Tamburini}},
  \bibinfo {author} {\bibfnamefont {F.}~\bibnamefont {Fiuza}}, \bibinfo
  {author} {\bibfnamefont {L.}~\bibnamefont {Gremillet}},\ and\ \bibinfo
  {author} {\bibfnamefont {S.}~\bibnamefont {Corde}},\ }\href
  {https://doi.org/10.1103/PhysRevResearch.4.023085} {\bibfield  {journal}
  {\bibinfo  {journal} {Phys. Rev. Res.}\ }\textbf {\bibinfo {volume} {4}},\
  \bibinfo {pages} {023085} (\bibinfo {year} {2022})}\BibitemShut {NoStop}%
\bibitem [{\citenamefont {Keinigs}\ and\ \citenamefont
  {Jones}(1987)}]{Keinings1987}%
  \BibitemOpen
  \bibfield  {author} {\bibinfo {author} {\bibfnamefont {R.}~\bibnamefont
  {Keinigs}}\ and\ \bibinfo {author} {\bibfnamefont {M.~E.}\ \bibnamefont
  {Jones}},\ }\href {https://doi.org/10.1063/1.866183} {\bibfield  {journal}
  {\bibinfo  {journal} {The Physics of Fluids}\ }\textbf {\bibinfo {volume}
  {30}},\ \bibinfo {pages} {252} (\bibinfo {year} {1987})},\ \Eprint
  {https://arxiv.org/abs/https://pubs.aip.org/aip/pfl/article-pdf/30/1/252/12365569/252\_1\_online.pdf}
  {https://pubs.aip.org/aip/pfl/article-pdf/30/1/252/12365569/252\_1\_online.pdf}
  \BibitemShut {NoStop}%
\bibitem [{\citenamefont {Katsouleas}\ \emph {et~al.}(1987)\citenamefont
  {Katsouleas}, \citenamefont {Wilks}, \citenamefont {Chen}, \citenamefont
  {Dawson},\ and\ \citenamefont {Su}}]{Katsouleas1987}%
  \BibitemOpen
  \bibfield  {author} {\bibinfo {author} {\bibfnamefont {T.~C.}\ \bibnamefont
  {Katsouleas}}, \bibinfo {author} {\bibfnamefont {S.}~\bibnamefont {Wilks}},
  \bibinfo {author} {\bibfnamefont {P.}~\bibnamefont {Chen}}, \bibinfo {author}
  {\bibfnamefont {J.~M.}\ \bibnamefont {Dawson}},\ and\ \bibinfo {author}
  {\bibfnamefont {J.~J.}\ \bibnamefont {Su}},\ }\href@noop {} {\bibfield
  {journal} {\bibinfo  {journal} {Part. Accel.}\ }\textbf {\bibinfo {volume}
  {22}},\ \bibinfo {pages} {81} (\bibinfo {year} {1987})}\BibitemShut {NoStop}%
\bibitem [{\citenamefont {Clayton}\ \emph {et~al.}(2016)\citenamefont
  {Clayton}, \citenamefont {Adli}, \citenamefont {Allen}, \citenamefont {An},
  \citenamefont {Clarke}, \citenamefont {Corde}, \citenamefont {Frederico},
  \citenamefont {Gessner}, \citenamefont {Green}, \citenamefont {Hogan},
  \citenamefont {Joshi}, \citenamefont {Litos}, \citenamefont {Lu},
  \citenamefont {Marsh}, \citenamefont {Mori}, \citenamefont
  {Vafaei-Najafabadi}, \citenamefont {Xu},\ and\ \citenamefont
  {Yakimenko}}]{Clayton2016}%
  \BibitemOpen
  \bibfield  {author} {\bibinfo {author} {\bibfnamefont {C.~E.}\ \bibnamefont
  {Clayton}}, \bibinfo {author} {\bibfnamefont {E.}~\bibnamefont {Adli}},
  \bibinfo {author} {\bibfnamefont {J.}~\bibnamefont {Allen}}, \bibinfo
  {author} {\bibfnamefont {W.}~\bibnamefont {An}}, \bibinfo {author}
  {\bibfnamefont {C.~I.}\ \bibnamefont {Clarke}}, \bibinfo {author}
  {\bibfnamefont {S.}~\bibnamefont {Corde}}, \bibinfo {author} {\bibfnamefont
  {J.}~\bibnamefont {Frederico}}, \bibinfo {author} {\bibfnamefont
  {S.}~\bibnamefont {Gessner}}, \bibinfo {author} {\bibfnamefont {S.~Z.}\
  \bibnamefont {Green}}, \bibinfo {author} {\bibfnamefont {M.~J.}\ \bibnamefont
  {Hogan}}, \bibinfo {author} {\bibfnamefont {C.}~\bibnamefont {Joshi}},
  \bibinfo {author} {\bibfnamefont {M.}~\bibnamefont {Litos}}, \bibinfo
  {author} {\bibfnamefont {W.}~\bibnamefont {Lu}}, \bibinfo {author}
  {\bibfnamefont {K.~A.}\ \bibnamefont {Marsh}}, \bibinfo {author}
  {\bibfnamefont {W.~B.}\ \bibnamefont {Mori}}, \bibinfo {author}
  {\bibfnamefont {N.}~\bibnamefont {Vafaei-Najafabadi}}, \bibinfo {author}
  {\bibfnamefont {X.}~\bibnamefont {Xu}},\ and\ \bibinfo {author}
  {\bibfnamefont {V.}~\bibnamefont {Yakimenko}},\ }\href
  {https://doi.org/10.1038/ncomms12483} {\bibfield  {journal} {\bibinfo
  {journal} {Nature Communications}\ }\textbf {\bibinfo {volume} {7}},\
  \bibinfo {pages} {12483} (\bibinfo {year} {2016})}\BibitemShut {NoStop}%
\bibitem [{\citenamefont {Albert}\ \emph {et~al.}(2021)\citenamefont {Albert},
  \citenamefont {Couprie}, \citenamefont {Debus}, \citenamefont {Downer},
  \citenamefont {Faure}, \citenamefont {Flacco}, \citenamefont {Gizzi},
  \citenamefont {Grismayer}, \citenamefont {Huebl}, \citenamefont {Joshi},
  \citenamefont {Labat}, \citenamefont {Leemans}, \citenamefont {Maier},
  \citenamefont {Mangles}, \citenamefont {Mason}, \citenamefont {Mathieu},
  \citenamefont {Muggli}, \citenamefont {Nishiuchi}, \citenamefont {Osterhoff},
  \citenamefont {Rajeev}, \citenamefont {Schramm}, \citenamefont {Schreiber},
  \citenamefont {Thomas}, \citenamefont {Vay}, \citenamefont {Vranic},\ and\
  \citenamefont {Zeil}}]{Albert2021}%
  \BibitemOpen
  \bibfield  {author} {\bibinfo {author} {\bibfnamefont {F.}~\bibnamefont
  {Albert}}, \bibinfo {author} {\bibfnamefont {M.~E.}\ \bibnamefont {Couprie}},
  \bibinfo {author} {\bibfnamefont {A.}~\bibnamefont {Debus}}, \bibinfo
  {author} {\bibfnamefont {M.~C.}\ \bibnamefont {Downer}}, \bibinfo {author}
  {\bibfnamefont {J.}~\bibnamefont {Faure}}, \bibinfo {author} {\bibfnamefont
  {A.}~\bibnamefont {Flacco}}, \bibinfo {author} {\bibfnamefont {L.~A.}\
  \bibnamefont {Gizzi}}, \bibinfo {author} {\bibfnamefont {T.}~\bibnamefont
  {Grismayer}}, \bibinfo {author} {\bibfnamefont {A.}~\bibnamefont {Huebl}},
  \bibinfo {author} {\bibfnamefont {C.}~\bibnamefont {Joshi}}, \bibinfo
  {author} {\bibfnamefont {M.}~\bibnamefont {Labat}}, \bibinfo {author}
  {\bibfnamefont {W.~P.}\ \bibnamefont {Leemans}}, \bibinfo {author}
  {\bibfnamefont {A.~R.}\ \bibnamefont {Maier}}, \bibinfo {author}
  {\bibfnamefont {S.~P.~D.}\ \bibnamefont {Mangles}}, \bibinfo {author}
  {\bibfnamefont {P.}~\bibnamefont {Mason}}, \bibinfo {author} {\bibfnamefont
  {F.}~\bibnamefont {Mathieu}}, \bibinfo {author} {\bibfnamefont
  {P.}~\bibnamefont {Muggli}}, \bibinfo {author} {\bibfnamefont
  {M.}~\bibnamefont {Nishiuchi}}, \bibinfo {author} {\bibfnamefont
  {J.}~\bibnamefont {Osterhoff}}, \bibinfo {author} {\bibfnamefont {P.~P.}\
  \bibnamefont {Rajeev}}, \bibinfo {author} {\bibfnamefont {U.}~\bibnamefont
  {Schramm}}, \bibinfo {author} {\bibfnamefont {J.}~\bibnamefont {Schreiber}},
  \bibinfo {author} {\bibfnamefont {A.~G.~R.}\ \bibnamefont {Thomas}}, \bibinfo
  {author} {\bibfnamefont {J.-L.}\ \bibnamefont {Vay}}, \bibinfo {author}
  {\bibfnamefont {M.}~\bibnamefont {Vranic}},\ and\ \bibinfo {author}
  {\bibfnamefont {K.}~\bibnamefont {Zeil}},\ }\href
  {https://doi.org/10.1088/1367-2630/abcc62} {\bibfield  {journal} {\bibinfo
  {journal} {New Journal of Physics}\ }\textbf {\bibinfo {volume} {23}},\
  \bibinfo {pages} {031101} (\bibinfo {year} {2021})}\BibitemShut {NoStop}%
\bibitem [{\citenamefont {Moreira}\ \emph {et~al.}(2023)\citenamefont
  {Moreira}, \citenamefont {Muggli},\ and\ \citenamefont
  {Vieira}}]{Moreira2023}%
  \BibitemOpen
  \bibfield  {author} {\bibinfo {author} {\bibfnamefont {M.}~\bibnamefont
  {Moreira}}, \bibinfo {author} {\bibfnamefont {P.}~\bibnamefont {Muggli}},\
  and\ \bibinfo {author} {\bibfnamefont {J.}~\bibnamefont {Vieira}},\ }\href
  {https://doi.org/10.1103/PhysRevLett.130.115001} {\bibfield  {journal}
  {\bibinfo  {journal} {Phys. Rev. Lett.}\ }\textbf {\bibinfo {volume} {130}},\
  \bibinfo {pages} {115001} (\bibinfo {year} {2023})}\BibitemShut {NoStop}%
\bibitem [{\citenamefont {Gschwendtner}\ \emph {et~al.}(2022)\citenamefont
  {Gschwendtner}, \citenamefont {Lotov}, \citenamefont {Muggli}, \citenamefont
  {Wing} \emph {et~al.}}]{Gschwendtner2022}%
  \BibitemOpen
  \bibfield  {author} {\bibinfo {author} {\bibfnamefont {E.}~\bibnamefont
  {Gschwendtner}}, \bibinfo {author} {\bibfnamefont {K.}~\bibnamefont {Lotov}},
  \bibinfo {author} {\bibfnamefont {P.}~\bibnamefont {Muggli}}, \bibinfo
  {author} {\bibfnamefont {M.}~\bibnamefont {Wing}}, \emph {et~al.} (\bibinfo
  {collaboration} {AWAKE Collaboration}),\ }\bibfield  {journal} {\bibinfo
  {journal} {Symmetry}\ }\textbf {\bibinfo {volume} {14}},\ \href
  {https://doi.org/10.3390/sym14081680} {10.3390/sym14081680} (\bibinfo {year}
  {2022})\BibitemShut {NoStop}%
\bibitem [{\citenamefont {Shukla}\ \emph {et~al.}(2018)\citenamefont {Shukla},
  \citenamefont {Vieira}, \citenamefont {Muggli}, \citenamefont {Sarri},
  \citenamefont {Fonseca},\ and\ \citenamefont {Silva}}]{Shukla2018}%
  \BibitemOpen
  \bibfield  {author} {\bibinfo {author} {\bibfnamefont {N.}~\bibnamefont
  {Shukla}}, \bibinfo {author} {\bibfnamefont {J.}~\bibnamefont {Vieira}},
  \bibinfo {author} {\bibfnamefont {P.}~\bibnamefont {Muggli}}, \bibinfo
  {author} {\bibfnamefont {G.}~\bibnamefont {Sarri}}, \bibinfo {author}
  {\bibfnamefont {R.}~\bibnamefont {Fonseca}},\ and\ \bibinfo {author}
  {\bibfnamefont {L.~O.}\ \bibnamefont {Silva}},\ }\href
  {https://doi.org/10.1017/S0022377818000314} {\bibfield  {journal} {\bibinfo
  {journal} {Journal of Plasma Physics}\ }\textbf {\bibinfo {volume} {84}},\
  \bibinfo {pages} {905840302} (\bibinfo {year} {2018})}\BibitemShut {NoStop}%
\bibitem [{\citenamefont {Pukhov}(2016)}]{Pukhov2016}%
  \BibitemOpen
  \bibfield  {author} {\bibinfo {author} {\bibfnamefont {A.}~\bibnamefont
  {Pukhov}},\ }\bibfield  {journal} {\bibinfo  {journal} {CERN Yellow Reports}\
  }\href {https://doi.org/10.5170/CERN-2016-001.181}
  {10.5170/CERN-2016-001.181} (\bibinfo {year} {2016})\BibitemShut {NoStop}%
\bibitem [{\citenamefont {Pukhov}(1999)}]{Pukhov1999}%
  \BibitemOpen
  \bibfield  {author} {\bibinfo {author} {\bibfnamefont {A.}~\bibnamefont
  {Pukhov}},\ }\href {https://doi.org/10.1017/S0022377899007515} {\bibfield
  {journal} {\bibinfo  {journal} {Journal of Plasma Physics}\ }\textbf
  {\bibinfo {volume} {61}},\ \bibinfo {pages} {425–433} (\bibinfo {year}
  {1999})}\BibitemShut {NoStop}%
\bibitem [{\citenamefont {Pukhov}\ \emph {et~al.}(2011)\citenamefont {Pukhov},
  \citenamefont {Kumar}, \citenamefont {T\"uckmantel}, \citenamefont
  {Upadhyay}, \citenamefont {Lotov}, \citenamefont {Muggli}, \citenamefont
  {Khudik}, \citenamefont {Siemon},\ and\ \citenamefont {Shvets}}]{Pukhov2011}%
  \BibitemOpen
  \bibfield  {author} {\bibinfo {author} {\bibfnamefont {A.}~\bibnamefont
  {Pukhov}}, \bibinfo {author} {\bibfnamefont {N.}~\bibnamefont {Kumar}},
  \bibinfo {author} {\bibfnamefont {T.}~\bibnamefont {T\"uckmantel}}, \bibinfo
  {author} {\bibfnamefont {A.}~\bibnamefont {Upadhyay}}, \bibinfo {author}
  {\bibfnamefont {K.}~\bibnamefont {Lotov}}, \bibinfo {author} {\bibfnamefont
  {P.}~\bibnamefont {Muggli}}, \bibinfo {author} {\bibfnamefont
  {V.}~\bibnamefont {Khudik}}, \bibinfo {author} {\bibfnamefont
  {C.}~\bibnamefont {Siemon}},\ and\ \bibinfo {author} {\bibfnamefont
  {G.}~\bibnamefont {Shvets}},\ }\href
  {https://doi.org/10.1103/PhysRevLett.107.145003} {\bibfield  {journal}
  {\bibinfo  {journal} {Phys. Rev. Lett.}\ }\textbf {\bibinfo {volume} {107}},\
  \bibinfo {pages} {145003} (\bibinfo {year} {2011})}\BibitemShut {NoStop}%
\bibitem [{\citenamefont {Graw}\ \emph {et~al.}(2022)\citenamefont {Graw},
  \citenamefont {Weidl},\ and\ \citenamefont {Jenko}}]{Graw2022}%
  \BibitemOpen
  \bibfield  {author} {\bibinfo {author} {\bibfnamefont {J.~M.}\ \bibnamefont
  {Graw}}, \bibinfo {author} {\bibfnamefont {M.~S.}\ \bibnamefont {Weidl}},\
  and\ \bibinfo {author} {\bibfnamefont {F.}~\bibnamefont {Jenko}},\ }\href
  {https://doi.org/10.3847/1538-4357/ac9bf1} {\bibfield  {journal} {\bibinfo
  {journal} {The Astrophysical Journal}\ }\textbf {\bibinfo {volume} {940}},\
  \bibinfo {pages} {172} (\bibinfo {year} {2022})}\BibitemShut {NoStop}%
\bibitem [{\citenamefont {Fonseca}\ \emph {et~al.}(2002)\citenamefont
  {Fonseca}, \citenamefont {Silva}, \citenamefont {Tsung}, \citenamefont
  {Decyk}, \citenamefont {Lu}, \citenamefont {Ren}, \citenamefont {Mori},
  \citenamefont {Deng}, \citenamefont {Lee}, \citenamefont {Katsouleas},\ and\
  \citenamefont {Adam}}]{Fonseca2002}%
  \BibitemOpen
  \bibinfo {editor} {\bibfnamefont {R.}~\bibnamefont {Fonseca}}, \bibinfo
  {editor} {\bibfnamefont {L.}~\bibnamefont {Silva}}, \bibinfo {editor}
  {\bibfnamefont {F.}~\bibnamefont {Tsung}}, \bibinfo {editor} {\bibfnamefont
  {V.}~\bibnamefont {Decyk}}, \bibinfo {editor} {\bibfnamefont
  {W.}~\bibnamefont {Lu}}, \bibinfo {editor} {\bibfnamefont {C.}~\bibnamefont
  {Ren}}, \bibinfo {editor} {\bibfnamefont {W.}~\bibnamefont {Mori}}, \bibinfo
  {editor} {\bibfnamefont {S.}~\bibnamefont {Deng}}, \bibinfo {editor}
  {\bibfnamefont {S.}~\bibnamefont {Lee}}, \bibinfo {editor} {\bibfnamefont
  {T.}~\bibnamefont {Katsouleas}},\ and\ \bibinfo {editor} {\bibfnamefont
  {J.}~\bibnamefont {Adam}},\ eds.,\ \href
  {https://doi.org/10.1007/3-540-47789-6_36} {\emph {\bibinfo {title} {OSIRIS:
  A Three-Dimensional, Fully Relativistic Particle in Cell Code for Modeling
  Plasma Based Accelerators}}},\ \bibinfo {series} {Computational Science -
  ICCS 2002}\ No.\ \bibinfo {number} {2331}\ (\bibinfo  {publisher} {Springer
  Berlin Heidelberg},\ \bibinfo {year} {2002})\BibitemShut {NoStop}%
\bibitem [{\citenamefont {Bret}\ and\ \citenamefont
  {Deutsch}(2006)}]{Bret2006}%
  \BibitemOpen
  \bibfield  {author} {\bibinfo {author} {\bibfnamefont {A.}~\bibnamefont
  {Bret}}\ and\ \bibinfo {author} {\bibfnamefont {C.}~\bibnamefont {Deutsch}},\
  }\href {https://doi.org/10.1063/1.2196876} {\bibfield  {journal} {\bibinfo
  {journal} {Physics of Plasmas}\ }\textbf {\bibinfo {volume} {13}},\ \bibinfo
  {pages} {042106} (\bibinfo {year} {2006})},\ \Eprint
  {https://arxiv.org/abs/https://pubs.aip.org/aip/pop/article-pdf/doi/10.1063/1.2196876/16011061/042106\_1\_online.pdf}
  {https://pubs.aip.org/aip/pop/article-pdf/doi/10.1063/1.2196876/16011061/042106\_1\_online.pdf}
  \BibitemShut {NoStop}%
\bibitem [{Note1()}]{Note1}%
  \BibitemOpen
  \bibinfo {note} {$400\protect \,\protect \mathrm {GeV}$ proton bunch with a
  total charge of $43\protect \,\protect \mathrm {nC}$, an rms width $\sigma _r
  = 0.5\protect \,\protect \mathrm {mm}$, a normalised emittance of
  $2.5\protect \,\protect \mathrm {mm}\protect \,\protect \mathrm {mrad}$, and
  a longitudinally Gaussian profile with $\sigma _\zeta /c = 163\protect
  \,\protect \mathrm {ps}$ ($\DOTSI \intop \ilimits@ \omega _\beta ^2\mathop
  {}\protect \!\protect \mathrm {d}\zeta =(2\pi )^{1/2}\omega _\beta ^2\sigma
  _\zeta $). The plasma length was $c\tau =10\protect \,\protect \mathrm
  {m}$}\BibitemShut {NoStop}%
\bibitem [{Note2()}]{Note2}%
  \BibitemOpen
  \bibinfo {note} {$0.06\protect \,\protect \mathrm {GeV}$ electron bunch with
  a total charge of $1\protect \,\protect \mathrm {nC}$, an rms width $\sigma
  _r = 0.065\protect \,\protect \mathrm {mm}$, a normalised emittance of
  $6\protect \,\protect \mathrm {mm}\protect \,\protect \mathrm {mrad}$, and an
  rms length $\sigma _\zeta /c = 5\protect \,\protect \mathrm {ps}$. The plasma
  length was $c\tau =0.02\protect \,\protect \mathrm {m}$}\BibitemShut
  {NoStop}%
\bibitem [{\citenamefont {Abramowitz}\ and\ \citenamefont
  {Stegun}(1964)}]{Abramowitz1964}%
  \BibitemOpen
  \bibfield  {author} {\bibinfo {author} {\bibfnamefont {M.}~\bibnamefont
  {Abramowitz}}\ and\ \bibinfo {author} {\bibfnamefont {I.~A.}\ \bibnamefont
  {Stegun}},\ }\href@noop {} {\emph {\bibinfo {title} {Handbook of Mathematical
  Functions with Formulas, Graphs, and Mathematical Tables}}},\ \bibinfo
  {edition} {ninth dover printing, tenth gpo printing}\ ed.\ (\bibinfo
  {publisher} {Dover},\ \bibinfo {address} {New York},\ \bibinfo {year}
  {1964})\BibitemShut {NoStop}%
\bibitem [{\citenamefont {Thode}(1976)}]{Thode1976}%
  \BibitemOpen
  \bibfield  {author} {\bibinfo {author} {\bibfnamefont {L.~E.}\ \bibnamefont
  {Thode}},\ }\href {https://doi.org/10.1063/1.861441} {\bibfield  {journal}
  {\bibinfo  {journal} {The Physics of Fluids}\ }\textbf {\bibinfo {volume}
  {19}},\ \bibinfo {pages} {305} (\bibinfo {year} {1976})},\ \Eprint
  {https://arxiv.org/abs/https://pubs.aip.org/aip/pfl/article-pdf/19/2/305/12638730/305\_1\_online.pdf}
  {https://pubs.aip.org/aip/pfl/article-pdf/19/2/305/12638730/305\_1\_online.pdf}
  \BibitemShut {NoStop}%
\bibitem [{\citenamefont {Califano}\ \emph {et~al.}(1998)\citenamefont
  {Califano}, \citenamefont {Prandi}, \citenamefont {Pegoraro},\ and\
  \citenamefont {Bulanov}}]{Califano1998}%
  \BibitemOpen
  \bibfield  {author} {\bibinfo {author} {\bibfnamefont {F.}~\bibnamefont
  {Califano}}, \bibinfo {author} {\bibfnamefont {R.}~\bibnamefont {Prandi}},
  \bibinfo {author} {\bibfnamefont {F.}~\bibnamefont {Pegoraro}},\ and\
  \bibinfo {author} {\bibfnamefont {S.~V.}\ \bibnamefont {Bulanov}},\ }\href
  {https://doi.org/10.1103/PhysRevE.58.7837} {\bibfield  {journal} {\bibinfo
  {journal} {Phys. Rev. E}\ }\textbf {\bibinfo {volume} {58}},\ \bibinfo
  {pages} {7837} (\bibinfo {year} {1998})}\BibitemShut {NoStop}%
\bibitem [{\citenamefont {Chang}\ \emph {et~al.}(2016)\citenamefont {Chang},
  \citenamefont {Broderick}, \citenamefont {Pfrommer}, \citenamefont
  {Puchwein}, \citenamefont {Lamberts}, \citenamefont {Shalaby},\ and\
  \citenamefont {Vasil}}]{Chang2016}%
  \BibitemOpen
  \bibfield  {author} {\bibinfo {author} {\bibfnamefont {P.}~\bibnamefont
  {Chang}}, \bibinfo {author} {\bibfnamefont {A.~E.}\ \bibnamefont
  {Broderick}}, \bibinfo {author} {\bibfnamefont {C.}~\bibnamefont {Pfrommer}},
  \bibinfo {author} {\bibfnamefont {E.}~\bibnamefont {Puchwein}}, \bibinfo
  {author} {\bibfnamefont {A.}~\bibnamefont {Lamberts}}, \bibinfo {author}
  {\bibfnamefont {M.}~\bibnamefont {Shalaby}},\ and\ \bibinfo {author}
  {\bibfnamefont {G.}~\bibnamefont {Vasil}},\ }\href
  {https://doi.org/10.3847/1538-4357/833/1/118} {\bibfield  {journal} {\bibinfo
   {journal} {The Astrophysical Journal}\ }\textbf {\bibinfo {volume} {833}},\
  \bibinfo {pages} {118} (\bibinfo {year} {2016})}\BibitemShut {NoStop}%
\bibitem [{\citenamefont {Bret}\ \emph {et~al.}(2008)\citenamefont {Bret},
  \citenamefont {Gremillet}, \citenamefont {B\'enisti},\ and\ \citenamefont
  {Lefebvre}}]{Bret2008}%
  \BibitemOpen
  \bibfield  {author} {\bibinfo {author} {\bibfnamefont {A.}~\bibnamefont
  {Bret}}, \bibinfo {author} {\bibfnamefont {L.}~\bibnamefont {Gremillet}},
  \bibinfo {author} {\bibfnamefont {D.}~\bibnamefont {B\'enisti}},\ and\
  \bibinfo {author} {\bibfnamefont {E.}~\bibnamefont {Lefebvre}},\ }\href
  {https://doi.org/10.1103/PhysRevLett.100.205008} {\bibfield  {journal}
  {\bibinfo  {journal} {Phys. Rev. Lett.}\ }\textbf {\bibinfo {volume} {100}},\
  \bibinfo {pages} {205008} (\bibinfo {year} {2008})}\BibitemShut {NoStop}%
\end{thebibliography}%

\end{document}